\documentclass[prl,twocolumn,superscriptaddress]{revtex4-1}
\usepackage{graphicx}
\usepackage{dcolumn}
\usepackage{bm}
\usepackage{color}
\usepackage{amsmath}
\usepackage{amsfonts}
\usepackage{soul}
\usepackage{url}
\usepackage{array,tabularx}
\usepackage{graphics}
\usepackage{times}
\usepackage{subfigure}
\usepackage[export]{adjustbox}

\begin{document}

\title{Non-Newtonian Topological Mechanical Metamaterials Using Feedback Control}
\author{Lea Sirota}\email{leabeilkin@tauex.tau.ac.il}
\affiliation{Raymond and Beverly Sackler School of Physics and Astronomy, Tel-Aviv University, Tel Aviv 69978, Israel}
\affiliation{School of Mechanical Engineering, Tel Aviv University, Tel Aviv 69978, Israel}
\author{Roni Ilan}
\affiliation{Raymond and Beverly Sackler School of Physics and Astronomy, Tel-Aviv University, Tel Aviv 69978, Israel}
\author{Yair Shokef}
\affiliation{School of Mechanical Engineering, Tel Aviv University, Tel Aviv 69978, Israel}
\affiliation{Sackler Center for Computational Molecular and Materials Science, Tel Aviv University, Tel Aviv 69978, Israel}
\affiliation{Center for Nonlinear Studies, Los Alamos National Laboratory, Los Alamos, New Mexico 87545, USA}
\affiliation{The Center for Physics and Chemistry of Living Systems, Tel Aviv University, Tel Aviv 69978, Israel}
\author{Yoav Lahini}
\affiliation{Raymond and Beverly Sackler School of Physics and Astronomy, Tel-Aviv University, Tel Aviv 69978, Israel}
\affiliation{The Center for Physics and Chemistry of Living Systems, Tel Aviv University, Tel Aviv 69978, Israel}

\begin{abstract}

We introduce a method to design topological mechanical metamaterials that are not constrained by Newtonian dynamics. 
The unit cells in a mechanical lattice are subjected to active feedback forces that are processed through autonomous controllers, pre-programmed to generate the desired local response in real-time. As an example, we focus on the quantum Haldane model, which is a two-band system with directional complex coupling terms, violating Newton's third law. We demonstrate that the required topological phase, characterized by chiral edge modes, can be achieved in an analogous mechanical system only with closed-loop control. We then show that our approach enables to realize, for the first time, a modified version of the Haldane model in a mechanical metamaterial. Here, the complex-valued couplings are polarized in a way that modes on opposite edges of a lattice propagate in the same direction, and are balanced by counter-propagating diffusive bulk modes. The proposed method is general and flexible, and could be used to realize arbitrary lattice parameters, such as non-local or nonlinear couplings, time dependent potentials, non-Hermitian dynamics, and more, on a single platform. 

\end{abstract}

\maketitle

The discovery of topologically protected wave phenomena in quantum physics \cite{thouless1982quantized,haldane1988model,kane2005quantum,bernevig2006quantum} with their exceptional immunity to back-scattering has recently inspired the search for realizations in classical systems, substituting the electronic band-structure with acoustic \cite{khanikaev2015topologically,he2016acoustic,zhang2017topological,yves2017topological} or photonic \cite{hafezi2013imaging,chen2014experimental,harari2018topological,bandres2018topological,ozawa2019topological} dispersion relations. 
These classical analogues are not merely a way to mimic well-known effects, but also a way to push the study of topological physics to new regimes \cite{rechtsman2013photonic,salerno2016floquet,fleury2016floquet,peng2016experimental,lee2019anatomy,scheibner2020non,rosa2020dynamics,brandenbourger2019non,helbig2020generalized}.
However, systems governed by Newtonian dynamics, such as mechanical structures supporting acoustic or elastic waves, usually do not naturally exhibit topological properties \cite{delplace2017topological}.
This insight is driving the current surge of activity aimed at designing topological mechanical metamaterials \cite{susstrunk2015observation,wang2015topological,nash2015topological,mousavi2015topologically,pal2017edge,chaunsali2018subwavelength,zhou2018quantum,liu2019design,ganti2020topological}.

The particular quantum effect considered for the classical realization plays a very important role in dictating the metamaterial design. As classical mechanical systems are constrained by Newtonian laws of motion, the range of known topological phenomena that can be observed is limited. 
The Quantum Spin Hall Effect \cite{kane2005quantum,bernevig2006quantum} or the Quantum Valley Hall Effect \cite{pan2014valley}, for which the spin-orbit coupling is  
obtained through breaking spatial symmetry in a lattice, comply with Newtonian dynamics, and can be implemented with purely passive components. Indeed, the vast majority of reports on mechanical topological metamaterials implement these effects, e.g. by designing the spacing of steel bars \cite{zhang2017topological} or bottle-like Helmholtz resonators \cite{yves2017topological} in an acoustic waveguide, the spacing of resonators on a plate \cite{chaunsali2018subwavelength}, the spring constants in a mass-spring lattice \cite{zhou2018quantum}, or a pendula array with intricate couplings \cite{susstrunk2015observation}, to name a few. 

Some quantum phases however, defy such a straightforward classical analogue. For example, creating an acoustic analogue of the quantum Hall effect, which requires breaking time reversal symmetry (TRS) (i.e. a Chern insulator), is considerably more involved, as passive design becomes insufficient. Consequently, there have only been a few reports of TRS breaking
in mechanical or acoustic systems; for fixed parameters and excitation frequency, a Chern insulator was emulated in a lattice of gyroscopes \cite{wang2015topological,nash2015topological}, and in a system of circulating fluids \cite{khanikaev2015topologically,souslov2017topological}. 
These realizations require auxiliary in-plane degrees of freedom (DOFs), i.e. actual physical in-plane rotation of masses or fluids.
Another method, based on Floquet acoustic crystals, employed temporal modulation of the acoustic parameters or the frequency \cite{salerno2016floquet,fleury2016floquet,peng2016experimental}. All these methods result in at least a four-band dispersion diagram.

It would be advantageous to break TRS with time-invariant parameters in a two-band metamaterial with out-of-plane DOFs alone, i.e. one DOF per mass. 
This is because it will enable reproduction of quantum effects that are associated with two-band systems \cite{haldane1988model,colomes2018antichiral,bhowmick2020antichiral,mannai2020strain,mandal2019antichiral,zhou2020observation}. In addition, it appears more feasible to realize such systems experimentally, as out-of-plane DOFs imply a scalar field, such as acoustic pressure or flexural waves.
However, when only out-of-plane DOFs (and time invariant parameters) are allowed, breaking TRS in acoustic or mechanical metamaterials requires lattice couplings that are inconsistent with the governing physical laws, including complex-valued directional, or non-reciprocal couplings, as discussed below. 

In this Letter we present a general method of realizing mechanical metamaterials that are not constrained by Newtonian laws of motion. 
At the heart of our method is the application of active forces to the masses in the out-of-plane direction, in real-time and autonomously, according to a pre-defined, programmable feedback control scheme.
To be concrete, we focus on a particular example and show how it is possible to break TRS using only out-of-plane DOFs, with a single DOF per site in a discrete mechanical lattice. 

Employing active control in the design of metamaterials has recently attracted considerable interest  \cite{chen2017hybrid,popa2015active,zangeneh2019active,sirota2019tunable,sirota2020active,darabi2020experimental,hofmann2019chiral,lee2019anatomy,scheibner2020non,rosa2020dynamics,brandenbourger2019non,helbig2020generalized}. 
In our system, an autonomous pre-programmed controller in each unit cell receives measurements of displacements and velocities of masses in neighboring lattice sites, processes them and feeds back to the control forces. The control operation therefore determines, in real-time, the dynamic response of the masses. Since the particular couplings are solely defined by the algorithm that we program into the controller, the feedback-based metamaterial is able to sustain any couplings (within hardware limitations), including those that are otherwise physically not achievable, such as directional or non-reciprocal couplings. 
Furthermore, a single system is not limited to emulate a particular quantum effect, but can be programmed to any other functionality.

We demonstrate our feedback-based design method by implementing, analytically and numerically, a classical mechanical analogue of the quantum Haldane model \cite{haldane1988model}, and the modified quantum Haldane model \cite{colomes2018antichiral,bhowmick2020antichiral,mannai2020strain,mandal2019antichiral,zhou2020observation}.
This is the first mechanical metamaterial realization of the quantum effect associated with the latter.
Both systems require non-Newtonian physics, which can be achieved only by the embedded feedback control mechanism. 
Below we present a detailed derivation of the controller that implements the Haldane model. For the modified Haldane model we show only the resulting dynamical simulations, with the details given in the supplemental file \cite{supplementary}.
To further demonstrate the versatility of our method, in \cite{supplementary} we also reprogram the embedded controller to realize a multipole pseudospin topological insulator on the same platform.

\begin{figure}[tb] 
\begin{center}
\begin{tabular}{l c}
\textbf{(a)} &  \includegraphics[width=6.8 cm]{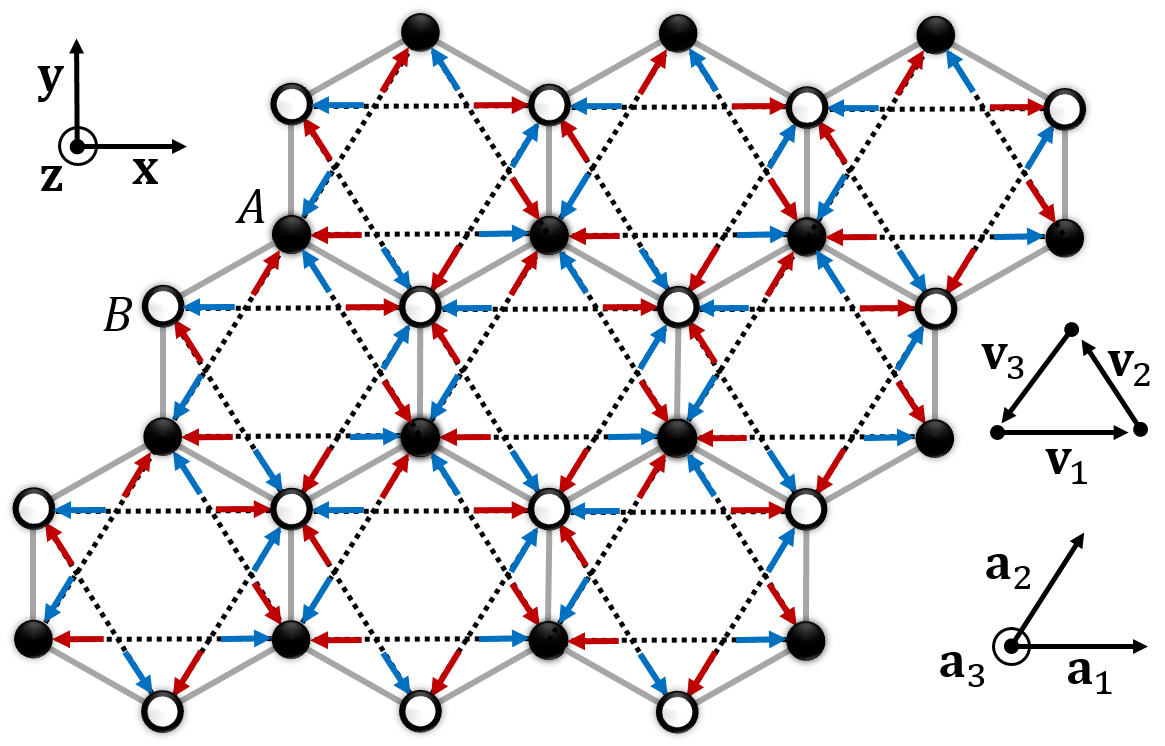} \\
 & \\
\textbf{(b)} & \includegraphics[width=6.2 cm]{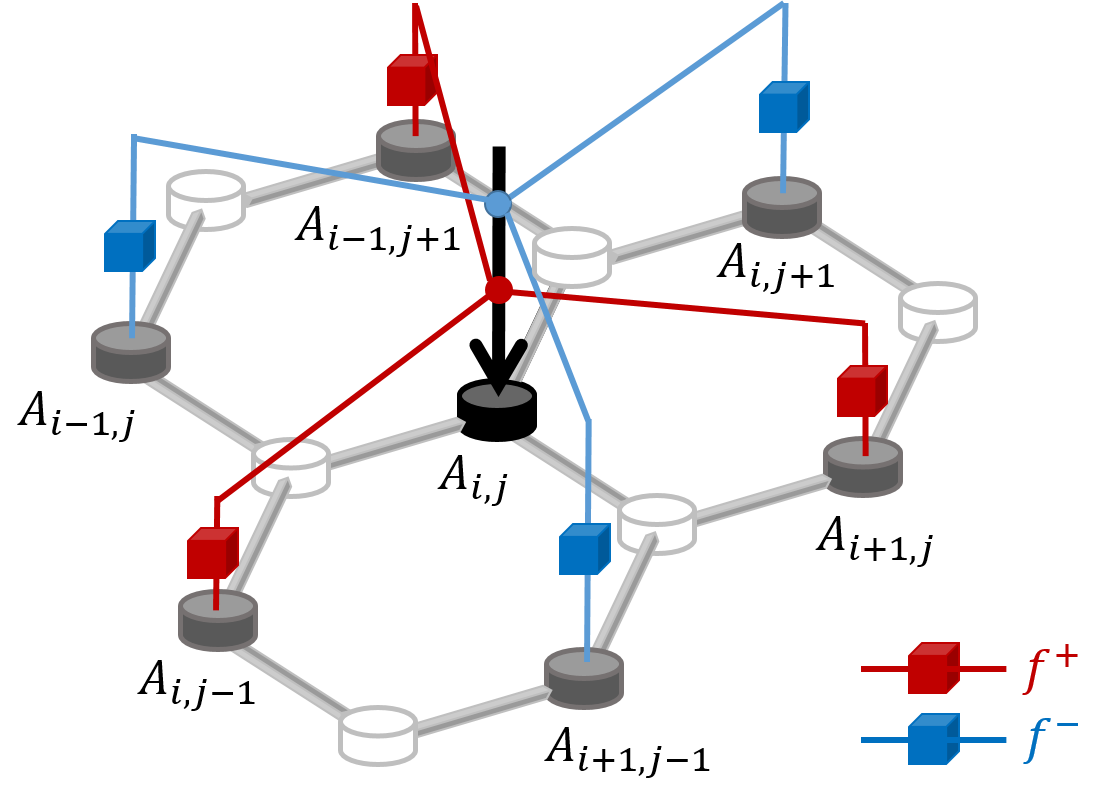} \\
 & \\
\textbf{(c)} & \includegraphics[width=6.0 cm]{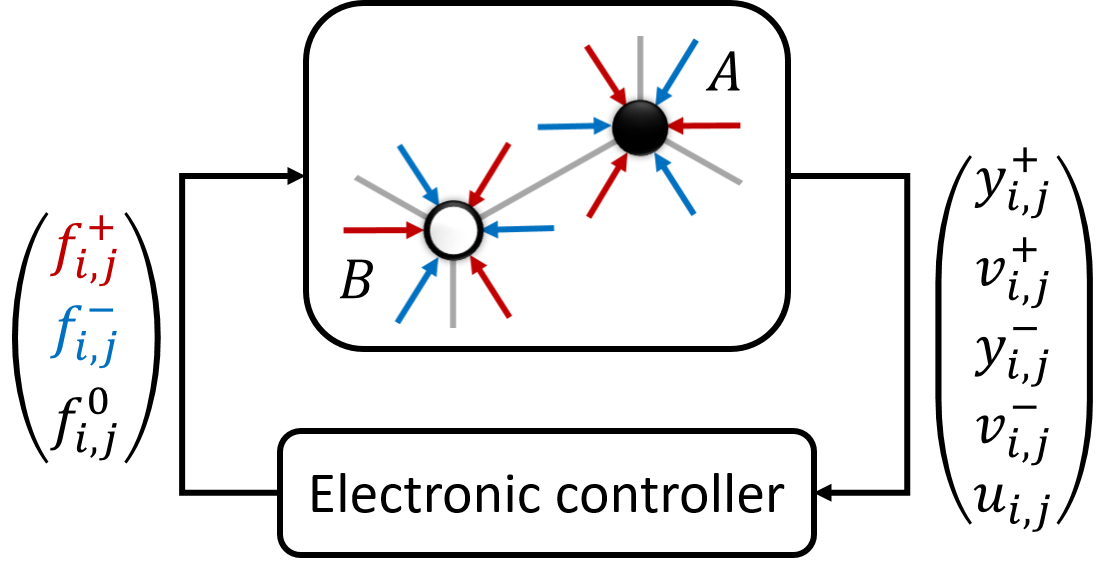} 
\end{tabular}
\caption{Feedback control scheme for generating the Haldane model in a mechanical metamaterial. (a) A honeycomb lattice in the $\{\textbf{a}_1,\textbf{a}_2\}$ space comprising identical masses (black and white circles) connected to nearest neighbors by Hookean springs (grey bars). The masses can move in the $\textbf{a}_3$ (out-of-plane) direction only, implying a single DOF per site. The red and blue arrows indicate out-of-plane displacement and velocity measurements of the next nearest neighbors (in $\{\textbf{v}_1,\textbf{v}_2,\textbf{v}_3\}$ directions). Dashed lines indicate the Haldane model bonds that are created in real-time when control is turned on. (b) The measurement scheme detailed for the $A$ site in the $\{i,j\}$ unit cell. The black arrow indicates the total control force $f^A=f^++f^-+f^0$. Red and blue arrows distinguish between measurements that are fed to controller gains $t_2e^{+i\phi}$ and $t_2e^{-i\phi}$ (red and blue cubes), applied through the $f^+$ and $f^-$ components, respectively. 
(c) Feedback control scheme of the entire $\{i,j\}$ unit cell, including all the measured signals.
}
\label{Scheme}
\end{center}    
\end{figure}

The Haldane model showed that the Quantum Hall Effect can be obtained without an external magnetic field, but rather by breaking TRS. It is defined on a honeycomb lattice spanned by $\{\textbf{a}_1,\textbf{a}_2\}$, which
consists of two interlacing triangular sub-lattices, exhibiting two sites per unit cell, $A$ and $B$, as illustrated in Fig. \ref{Scheme}(a). We denote the lattice constant by $a$. In our classical mechanical analogue, the circles are identical masses $m_0$ that can vibrate only along the vertical axis $\textbf{a}_3$. The grey bars indicate nearest neighbor couplings, which are equivalent to Hookean springs of stiffness $t_1>0$ connecting the masses. 
When only the $t_1$ springs exist, the lattice is analogous to graphene. 

The quantum Haldane model assumes additional next nearest neighbor bonds of a complex strength $t_2e^{\pm i\phi}$ in the directions $\textbf{v}_1,\textbf{v}_2,\textbf{v}_3$, indicated by the dashed black lines in Fig. \ref{Scheme}(a). In a mechanical context, such a bond represents a non-reciprocal coupling, $t_2e^{+i\phi}$ (red arrow) towards one mass and $t_2e^{-i\phi}$ (blue arrow) towards the other connected mass, which violates Newton's third law and is therefore non-physical. These couplings cannot be implemented with passive devices such as springs, lever arms etc., or with auxiliary active devices like gyroscopes that rely on in-plane DOFs. 
We realize these couplings using active closed-loop control.
Contrary to usual expectations, the complex values of the forces are physical in the time-harmonic regime because they are related to velocities rather than to displacements.
The full form of the Haldane model is captured by the Bloch Hamiltonian 
$H(\textbf{k})=\sum_{l=0}^3\mathcal{H}_l(\textbf{k})\sigma_l$, where $\textbf{k}$ is the wave vector, $\sigma_l$ are the Pauli matrices, and
\begin{equation}    \label{eq:H}
\begin{split}
&\mathcal{H}_0=\beta+2t_2\cos\phi\textstyle{\sum_{m=1}^3}\cos(\textbf{k}\cdot\textbf{v}_m),\\
&\mathcal{H}_1=-t_1\left(1+\cos(\textbf{k}\cdot\textbf{a}_1)+\cos(\textbf{k}\cdot\textbf{a}_2)\right),\\
&\mathcal{H}_2=t_1\left(\sin(\textbf{k}\cdot\textbf{a}_1)+\sin(\textbf{k}\cdot\textbf{a}_2)\right),\\
&\mathcal{H}_3=M-2t_2\sin\phi\textstyle{\sum_{m=1}^3}\sin(\textbf{k}\cdot\textbf{v}_m).
\end{split}
\end{equation}
In a quantum system, $\beta=0$. In the classical-mechanical analogue of graphene $\beta=3t_1$, indicating the restoring $t_1$ force from the three nearest neighbor springs, and is not related to Haldane's next nearest neighbor bonds. The constant $M$ accounts for a possible spatial inversion symmetry breaking, in addition to the TRS breaking provided by the $t_2$ bonds. 

The goal of our embedded control system is to create a classical-mechanical metamaterial, whose dynamics in lattice momentum space $\textbf{k}$ is given by
\begin{equation}  \label{eq:EQ}
\omega^2\textbf{p}(\textbf{k})=H(\textbf{k})\textbf{p}(\textbf{k}).
\end{equation}
Here $H(\textbf{k})$ is the Bloch Hamiltonian of the Haldane model, and $\textbf{p}(\textbf{k})$ is the complex amplitude vector of the $A$ and $B$ sites in momentum space. 
Our starting point is the graphene-like lattice ($t_1$ springs only), in which we denote the DOFs of each $\{i,j\}$ unit cell by $\textbf{u}_{i,j}(t)=(\begin{array}{c c}u^A_{i,j}(t), & u^B_{i,j}(t)\end{array})^T$.
The time domain unit cell dynamics, including external mechanical control forces $f^A_{i,j},f^B_{i,j}$ that are applied to the masses in the $a_3$ direction, reads
\begin{equation}   \label{eq:Motion_OL}
\begin{split}
&\ddot{u}^A_{i,j}=-3t_1u^A_{i,j}+t_1\left(u^B_{i,j}+u^B_{i+1,j}+u^B_{i,j+1}\right)+f^A_{i,j}, \\
&\ddot{u}^B_{i,j}=-3t_1u^B_{i,j}+t_1\left(u^A_{i,j}+u^A_{i-1,j}+u^A_{i,j-1}\right)+f^B_{i,j}.
\end{split}
\end{equation}
The control forces are decomposed into $f^A_{i,j}=f^{A+}_{i,j}+f^{A-}_{i,j}+f^{A0}_{i,j}$ and $f^B_{i,j}=f^{B+}_{i,j}+f^{B-}_{i,j}+f^{B0}_{i,j}$. The $f^+$ and $f^-$ components are responsible for generating the $t_2e^{+i\phi}$ and the $t_2e^{-i\phi}$ couplings, respectively, and $f^0$ is responsible for generating $M$.
As depicted in Fig. \ref{Scheme}(b), e.g. for the $A$ site, the $f^{A+}$ and $f^{A-}$ components receive measured signals of displacements and velocities of the $u^A_{i+1,j},u^A_{i-1,j+1},u^A_{i,j-1}$ and $u^A_{i-1,j},u^A_{i+1,j-1},u^A_{i,j+1}$ DOFs, as indicated by the red and blue arrows, respectively. These arrows are also shown on the multi-cell lattice segment in Fig. \ref{Scheme}(a). 
The $f^{A0}$ component is not depicted.
The measurements are processed in real-time by corresponding controllers, indicated by red and blue cubes. The control action is illustrated in Fig. \ref{Scheme}(c) for the $\{i,j\}$ unit cell. 
For each site $A$ and $B$ (the superscripts are omitted in the following), the control forces are related to the measured signals as
\begin{equation}    \label{eq:C_law}
\left( \begin{array}{c c c} f^+_{i,j} & f^-_{i,j} & f^0_{i,j} \end{array} \right)^T=C \left( \begin{array}{c c c c c} y^+_{i,j} & v^+_{i,j} & y^-_{i,j} & v^-_{i,j} & u_{i,j}\end{array} \right)^T,
\end{equation} 
where, for the $A$ site,
\begin{equation}    \label{eq:Measurements}
\begin{split}
&y^\pm _{i,j}=u_{i\pm 1,j}+u_{i\mp 1,j\pm 1}+u_{i,j\mp 1}, \\
&v^\pm _{i,j}=\dot{u}_{i\pm 1,j}+\dot{u}_{i\mp 1,j\pm 1}+\dot{u}_{i,j\mp 1}.
\end{split}
\end{equation}
For the $B$ site, the definitions of $y^+(v^+)$ and $y^-(v^-)$ in \eqref{eq:Measurements} are swapped. The control matrix $C$ at each $\{i,j\}$ unit cell, for both $A$ and $B$ sites, is given by
\begin{equation}    \label{eq:C}
C=\left( \begin{array}{ccccc} t_2\cos \phi & \frac{t_2}{\omega}\sin \phi & 0 & 0 & 0 \\ 0 & 0 & t_2\cos \phi & -\frac{t_2}{\omega}\sin \phi & 0 \\ 0 & 0 & 0 & 0 & \pm M \end{array} \right).
\end{equation}
The sign of $M$ in \eqref{eq:C} is positive (negative) for the $A$ ($B$) sites. 
Since in frequency domain, the velocity $v$ is related to the displacement $u$ as $v=i\omega u$, the controller gains that generate the velocity couplings are normalized by the frequency of the source signal, which is a fixed scalar in a given working regime.
This normalization guarantees the frequency independence of the complex-valued non-reciprocal next nearest neighbor couplings required by the Haldane model Hamiltonian \eqref{eq:H}, which are created by the control forces in real-time.
The resulting closed-loop system is dynamically stable, and the control forces do not exceed the source force amplitude. 

A two-band system implies a scalar dynamical field, which significantly reduces the complexity of experimental realization compared to systems with higher number of bands. 
One possibility is implementation in an actual discrete mechanical system. The out-of-plane displacement may then be achieved by constraining weights, horizontally connected to nearest neighbors by prestressed harmonic springs, to move on vertical shafts through linear bearings. The spring constant needs to be tuned carefully to balance between friction reduction and dominance over gravity. 
An alternative implementation is by an acoustic pressure field, created in a two-dimensional waveguide by an array of loudspeakers. In both scenarios, the feedback control system in \eqref{eq:Motion_OL}-\eqref{eq:C} is realized by an autonomous micro-controller that processes the measurements at corresponding next nearest neighbor locations. 

Next we demonstrate that the mechanical system \eqref{eq:EQ} governed by the classical analogue of the Haldane model Hamiltonian \eqref{eq:H}, which we created with the control system \eqref{eq:C_law}-\eqref{eq:C}, reproduces all the known dynamic properties of the quantum Haldane model. 
This is not obvious, since the complex-valued couplings are retained only upon reaching the time-harmonic regime.
\begin{figure}[tb] 
\begin{center}
\begin{tabular}{l l}
\textbf{(a)} & \textbf{(b)} \\
\includegraphics[width=4.2 cm]{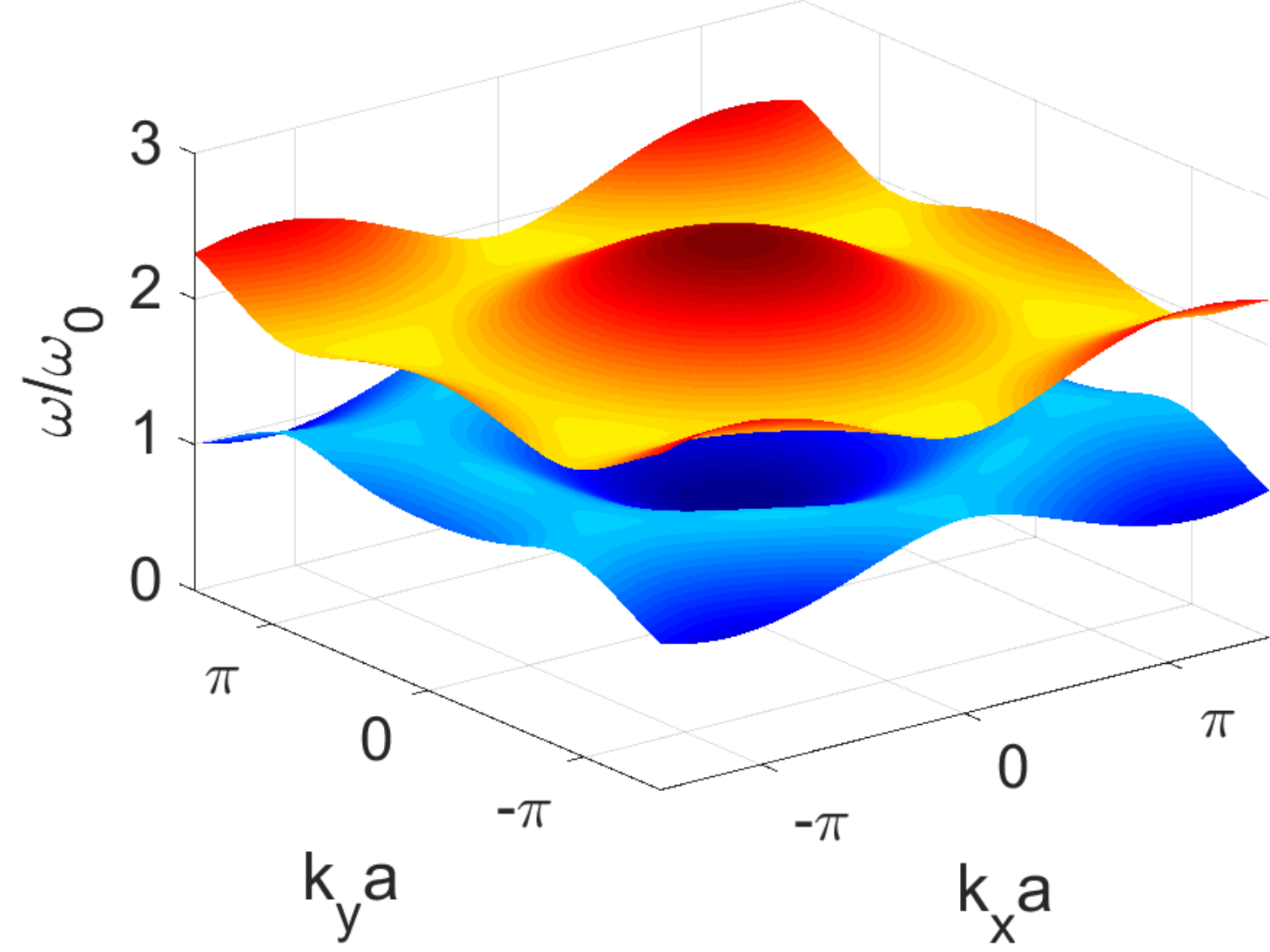} &
\includegraphics[width=4.1 cm]{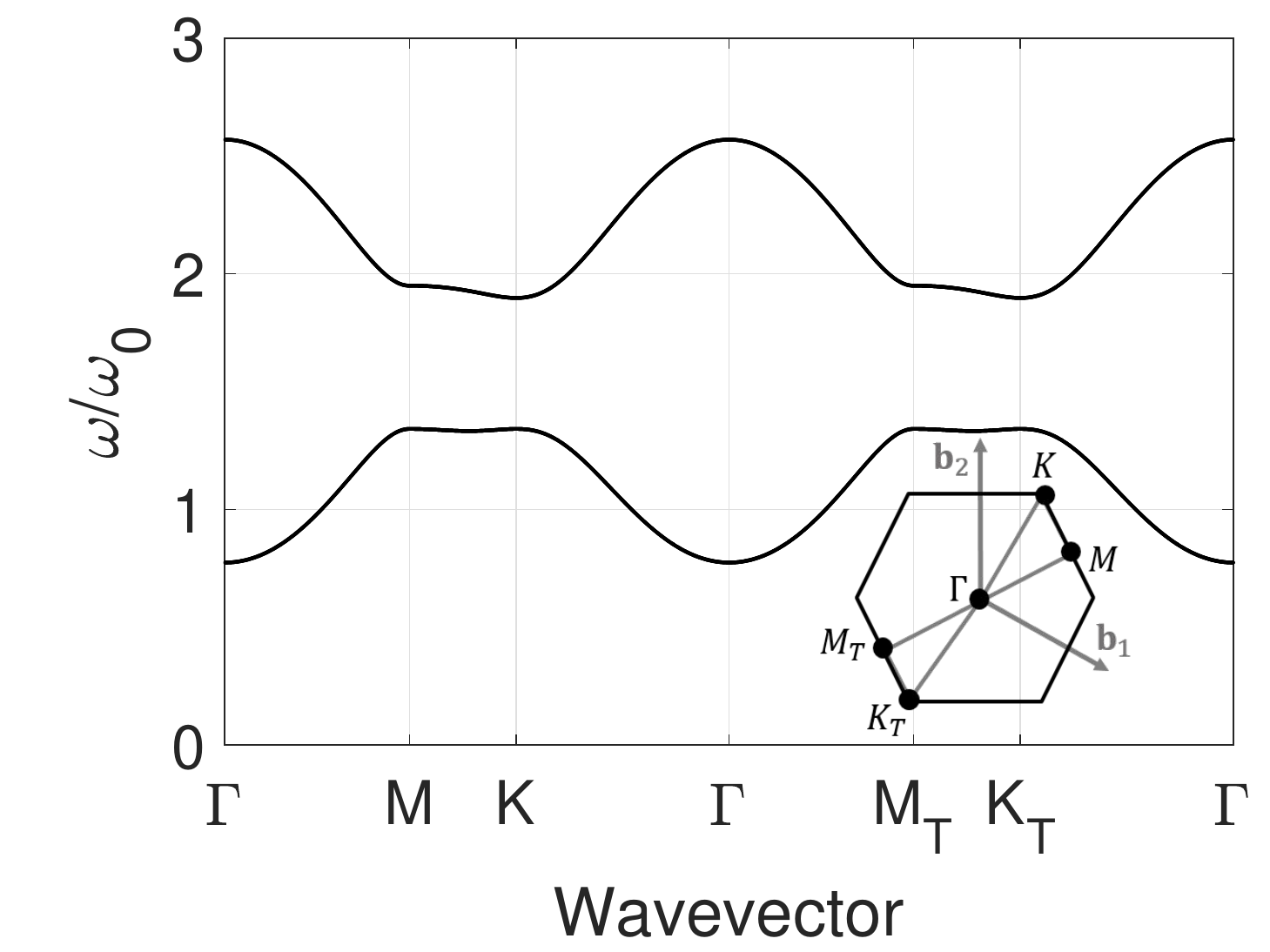} 
\end{tabular}
\caption{Infinite lattice dispersion relations of the Haldane model mechanical analogue. Dispersion diagram with parameters $\{t_2=0.2t_1,\phi=\pi/3,M=0\}$ is plotted over the entire Brillouin zone (a) and through the high symmetry points (b).}
\label{BulkDisp}
\end{center}    
\end{figure}
The first property that we analyze is the band-structure, here the acoustic dispersion, of an infinite lattice. 
Since \eqref{eq:EQ} represents a classical-mechanical system,
the eigenvalues are squared frequencies. 
The frequencies are kept real and positive due to the constant shift of the dispersion curves by the addition of $\beta=3t_1$ to the $\sigma_0$ term in \eqref{eq:H}. Since this addition does not change the eigenvectors, the metamaterial preserves the topological properties of the original quantum Haldane model.
We consider, for example, $\{\phi=\pi/3,M=0,t_2=0.2t_1\}$, which falls within the non-trivial topological regime of Chern number $n=+1$, according to the phase diagram of the quantum Haldane model \cite{haldane1988model}. 
The corresponding band-structure is depicted in Fig. \ref{BulkDisp}. 
The frequency scale is normalized by $\omega_0=\sqrt{t_1/m_0}$. See \cite{supplementary} for a different set of parameters, and for the actual parameter scale.
Similarly to the quantum system, the band-structure is not symmetric between the $\Gamma-M-K-\Gamma$ and the $\Gamma-M_T-K_T-\Gamma$ trajectories. 

\begin{figure}[tb] 
\begin{center}
\begin{tabular}{l l @{\hspace{0.1pt}} l @{\hspace{0.1pt}} l}
\textbf{(a)} & \textbf{(b)} & & \textbf{(c)} \\
\includegraphics[height=4.6 cm]{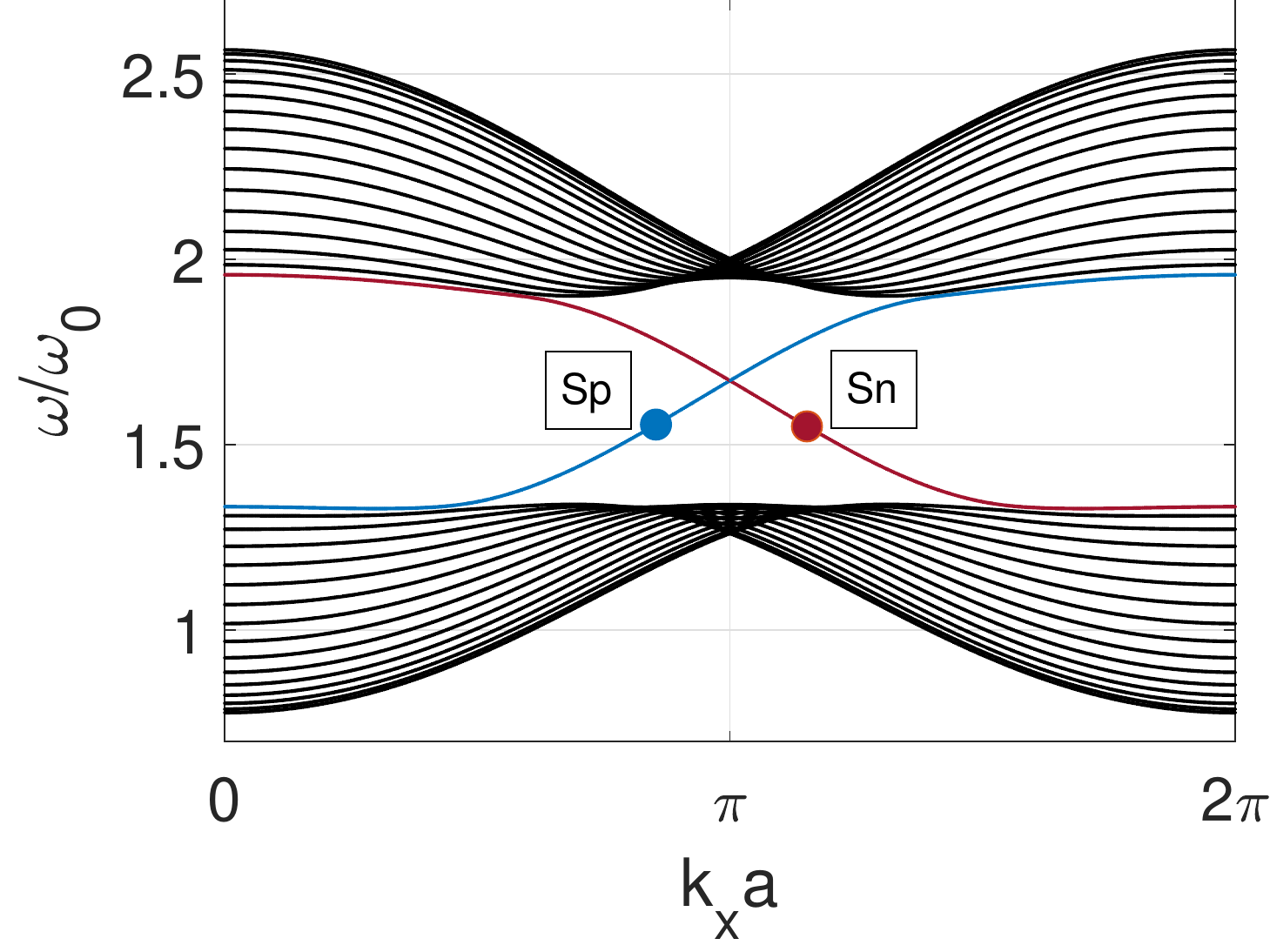} & \includegraphics[height=4.3 cm]{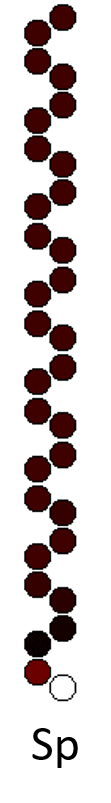} & \includegraphics[height=2.2 cm]{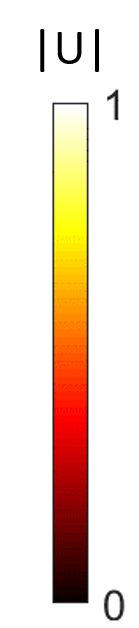} & \includegraphics[height=4.3 cm]{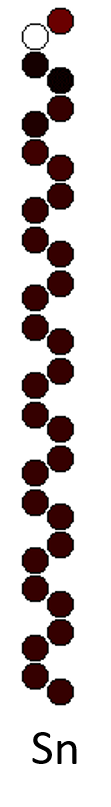}
\end{tabular}
\caption{Edge modes of the mechanical Haldane model. (a) Dispersion diagram of a lattice infinite in the $\textbf{x}$ direction and of eight honeycomb cells in the $\textbf{y}$ direction. Black lines indicate bulk states. The red (blue) line indicates top (bottom) edge state with negative (positive) group velocity. (b),(c) Corresponding eigenmodes of top and bottom edge states.}
\label{EdgeModeDisp}
\end{center}    
\end{figure}

Next we verify the reproduction of the quantum edge mode dispersion. Now our lattice is infinite in $\textbf{x}$ but finite in the $\textbf{y}$ direction. The dispersion diagram of an eight honeycomb cells strip is plotted in Fig. \ref{EdgeModeDisp}(a). 
As expected for the Haldane model, a state emerges inside the bulk bandgap, corresponding to top edge propagation with negative group velocity ($S_n$ point), and to bottom edge propagation with positive group velocity ($S_p$ point). The eigenmodes associated with these counter-propagating chiral edge states are depicted in Fig. \ref{EdgeModeDisp}(b),(c). The wave localization on the lattice edge is guaranteed by the topological property of the band-structure.  

Recently, a modified version of the Haldane model was proposed \cite{colomes2018antichiral,mandal2019antichiral,bhowmick2020antichiral,mannai2020strain}. Here both top and bottom edge modes propagate in the same direction, compensated by bulk modes that propagate in the opposite direction. Such anti-chiral edge states can exist in two-dimensional lattices if the bulk band-structure is gapless, as the number of left and right moving modes in a finite system must be the same.
A significant suppression of backscattering is provided for the edge modes due to their spatial separation from the bulk modes, whereas the bulk modes diffuse across the lattice width.
To obtain the modified Haldane model, one needs to flip the direction of the complex-valued next nearest neighbor couplings of one of the unit cell sites, e.g. of the $B$ site in Fig. \ref{Scheme}(a). 
The resulting mechanical analogue is a two-band non-Newtonian system, similarly to the original Haldane model, and can be realized in mechanical systems only with a feedback mechanism. The finite lattice band-structure and the corresponding controller are given in \cite{supplementary}. 

\begin{figure}[tb]
\begin{center}
\begin{tabular}{l l}
\textbf{(a)} Program $1$, $t=T_1$. & \textbf{(b)} Program $1$, $t=T_2$. \\
\includegraphics[width=4.1 cm]{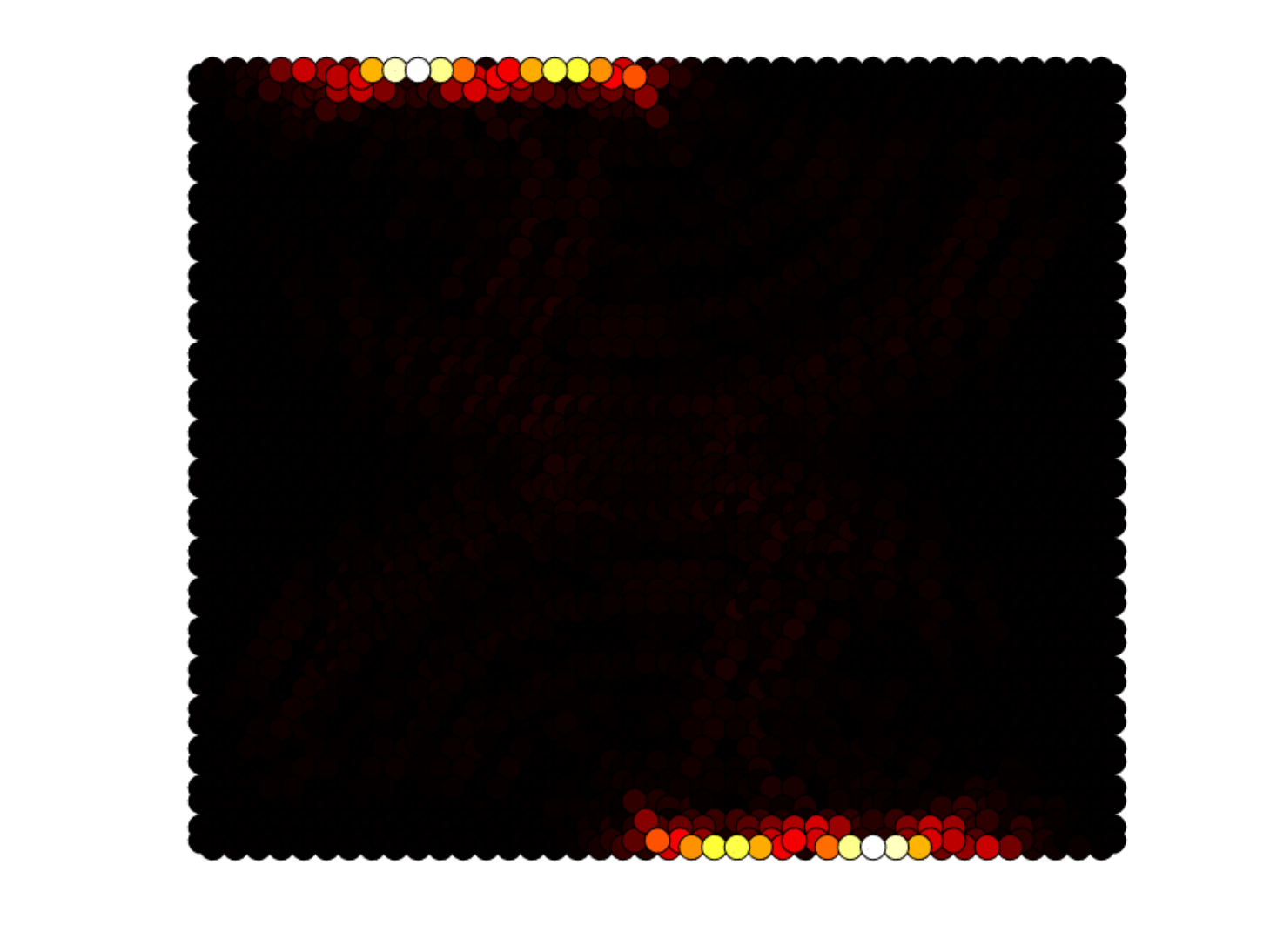} & \includegraphics[width=4.1 cm]{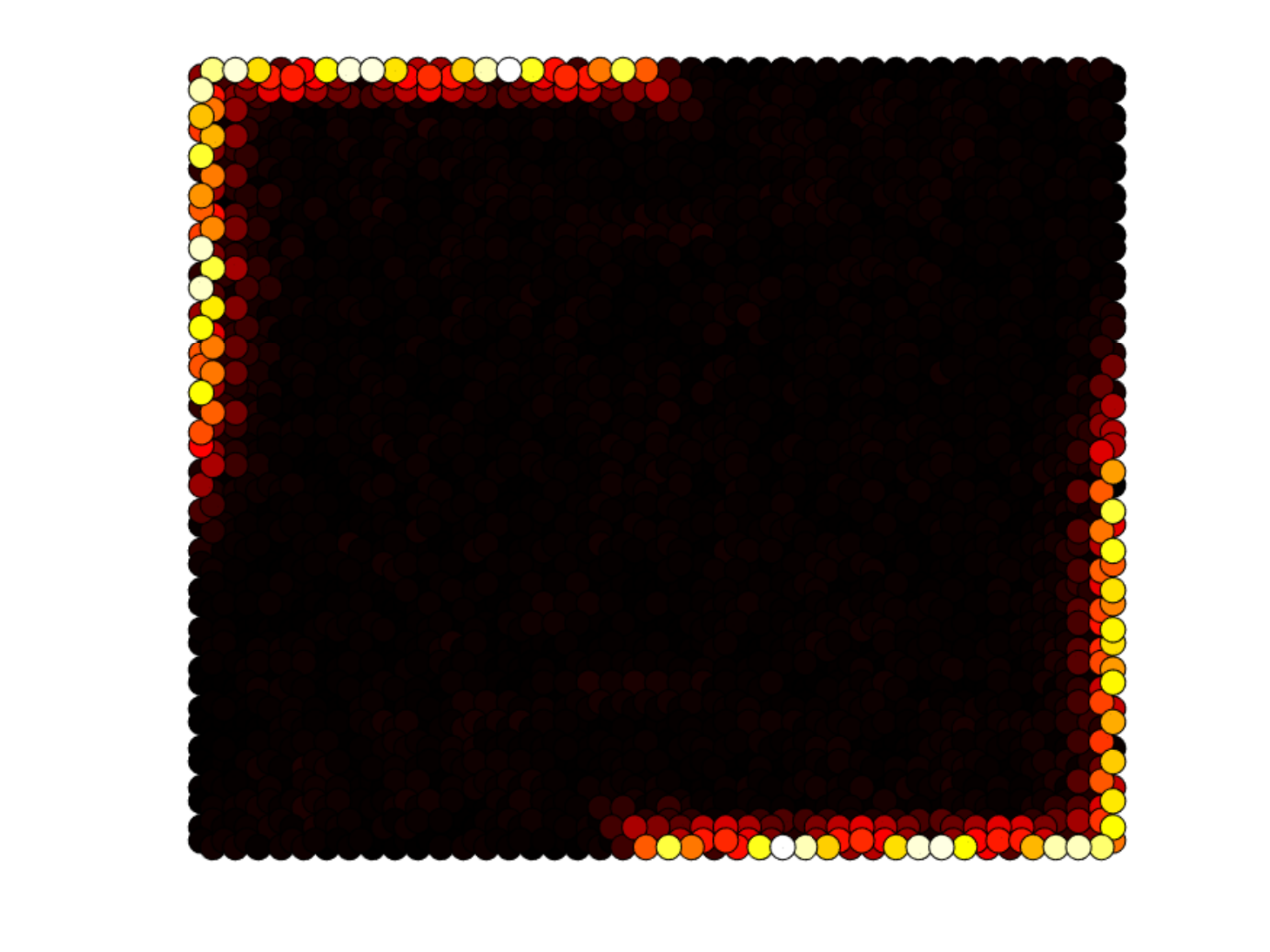} \\ 
\textbf{(c)} Program $2$, $t=T_1$. & \textbf{(d)} Program $2$, $t=T_2$. \\
\includegraphics[width=4.1 cm]{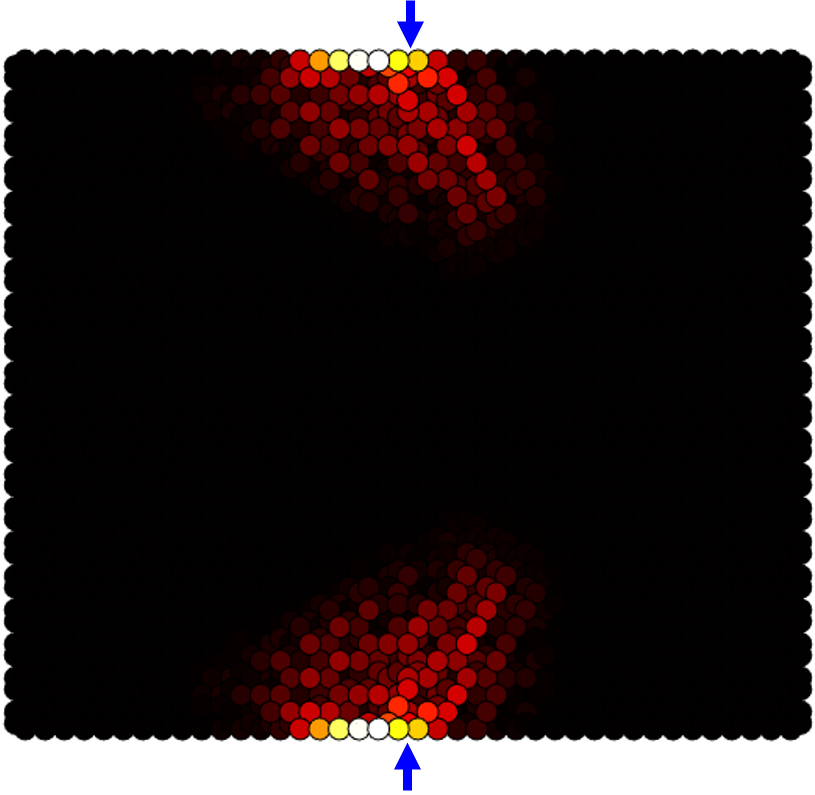}  & \includegraphics[width=4.1 cm]{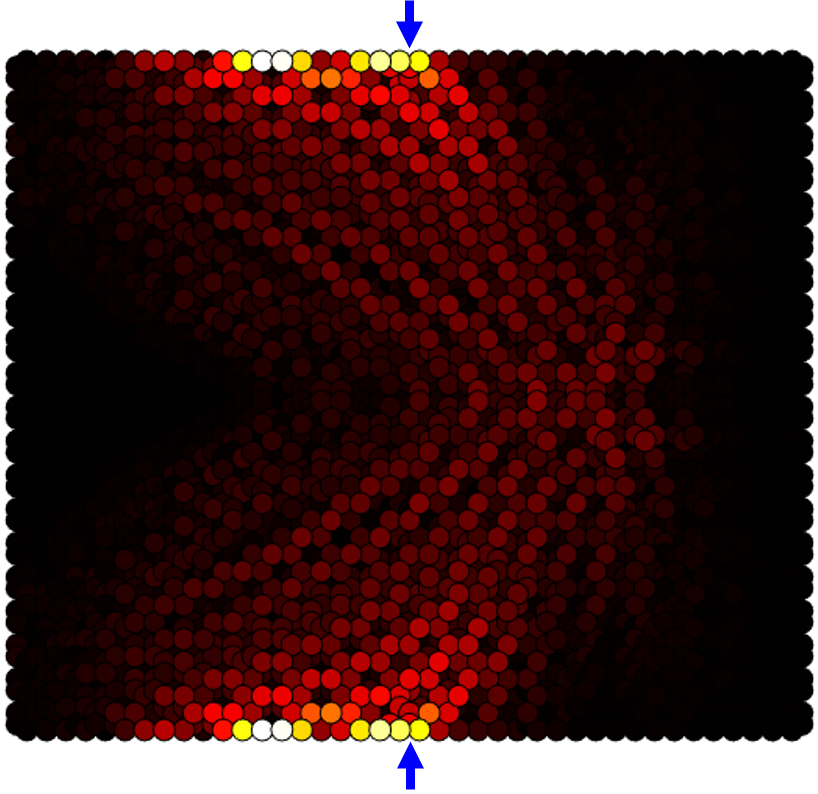}
\end{tabular}
\includegraphics[width=4.1 cm]{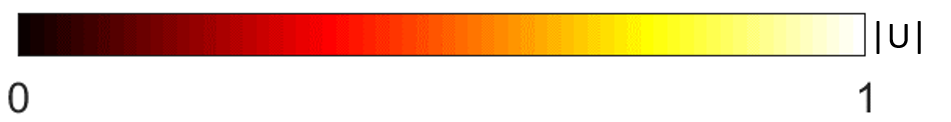}
\end{center}
\caption{Dynamical simulation of the feedback-based mechanical metamaterial. A force $F(t)=F_0e^{i\omega t}$, indicated by the blue arrows, is applied to a $20\times 40$ honeycomb lattice in the $\textbf{a}_3$ direction. The displacement responses of the masses, in $\textbf{a}_3$, are depicted at time instances $T_1<T_2$ (left and right columns). \textbf{(a)},\textbf{(b)} - Control program $1$, realizing the Haldane model. \textbf{(c)},\textbf{(d)} - Control program $2$, realizing the modified Haldane model.}
\label{Sim}   
\end{figure}

We now demonstrate that our metamaterial supports uni-directional edge wave propagation, as expected for both the original and the modified Haldane models. 
We perform two dynamical simulations of a finite size metamaterial ($20\times 40$ honeycomb net), which is operated in a real-time feedback loop, see Fig. \ref{Sim}.
Fixed boundary conditions along all edges are assumed, and the actuation frequency is set to $\omega=1.55\omega_0$. 
The system is excited by a time-harmonic force $F(t)=F_0e^{i\omega t}$ in the $\textbf{a}_3$ direction at the middle of the top and bottom edges, as indicated by the blue arrows in the figure. 
Closed-loop time responses of the masses out-of-plane displacements $u_{i,j}(t)$ (normalized by $F_0$) are shown at two time instances, $T_1<T_2$. At these times the control transients converged, and the system reached its dynamical steady state.

Figure \ref{Sim} (a,b) corresponds to control program 1, creating the Haldane model \eqref{eq:H}, according to \eqref{eq:EQ}-\eqref{eq:C} with $\{t_2=0.2t_1,\phi=\pi/3,M=0\}$.
Here the actuation frequency lies inside the bulk bandgap.
One clearly sees that at the top (bottom) edge the wave propagates to the left (right) corresponding to the $S_n$ ($S_p$) point in Fig. \ref{EdgeModeDisp}(a). Due to topological protection, the wave circumvents the sharp lattice corners without any back-scattering. 
Figure \ref{Sim} (c,d) corresponds to control program 2, creating the modified Haldane model with $\{t_2=0.1t_1,\phi=\pi/3,M=0\}$. As expected, both edge waves propagate to the left, and simultaneously, there are two bulk waves, of a considerably reduced intensity, that are swept to the right. 

To summarize, we proposed and analyzed a general feedback-based method for realizing topological mechanical metamaterials with arbitrary responses that are not constrained by Newton’s laws of motion. 
As an example, we implemented two topological two-band systems, analogous to the quantum Haldane, and the modified Haldane model, in a classical mechanical metamaterial. 
The required non-Newtonian complex-valued directional couplings between masses were generated via autonomous real-time control. 
We demonstrated that the resulting systems have all the properties of the quantum models, and support the expected wave propagation along the metamaterial edges. 
The method is general, and could be programmed to implement a wide variety of topological models in classical systems, relying both on Newtonian and non-Newtonian dynamics. 
To stress this point, in \cite{supplementary} we design a pseudospin multipole topological insulator on the same hardware platform, yet with a different feedback software.
The resulting system mimics the quantum spin Hall effect \cite{chaunsali2018subwavelength}, but without any spinning elements, using out-of-plane DOFs alone.
 
\textit{We thank Moshe Goldstein and Daniel Sabsovich for fruitful discussions. This research was supported in part by the Israel Science Foundation Grants No. 968/16 and 2096/18, by the Israeli Ministry of Science and Technology Grant No. 3-15671, by the US-Israel Binational Science Foundatio Grant No. 2018226, and by the National Science Foundation Grant No. NSF PHY-1748958.}

\bibliography{HaldaneModelbib}

\renewcommand{\thefigure}{S\arabic{figure}}
\renewcommand{\theequation}{S\arabic{equation}}
\setcounter{figure}{0}
\setcounter{equation}{0}

\clearpage

\begin{widetext}

\section{Supplemental Material}

\subsection{Haldane model - infinite system band-structure}

Figure S1 verifies that our feedback-based mechanical metamaterial reproduces the expected infinite-system dispersion relation of the quantum Haldane model for a different set of parameters than in the main text, specifically for $\{t_2=0.2t_1,\phi=\pi/3,M=3\sqrt{3}t_2\sin\phi\}$. This case falls exactly on the border between the topologically trivial ($n=0$) and nontrivial ($n=+1$) regimes, where $n$ indicates the Chern number. Due to TRS breaking, a gap is opened at the $K$ point but not at the $K_T$ point.

\begin{figure}[htpb] 
\begin{center}
\begin{tabular}{l l l l}
\textbf{(a)} & \includegraphics[width=5.4 cm, valign=t]{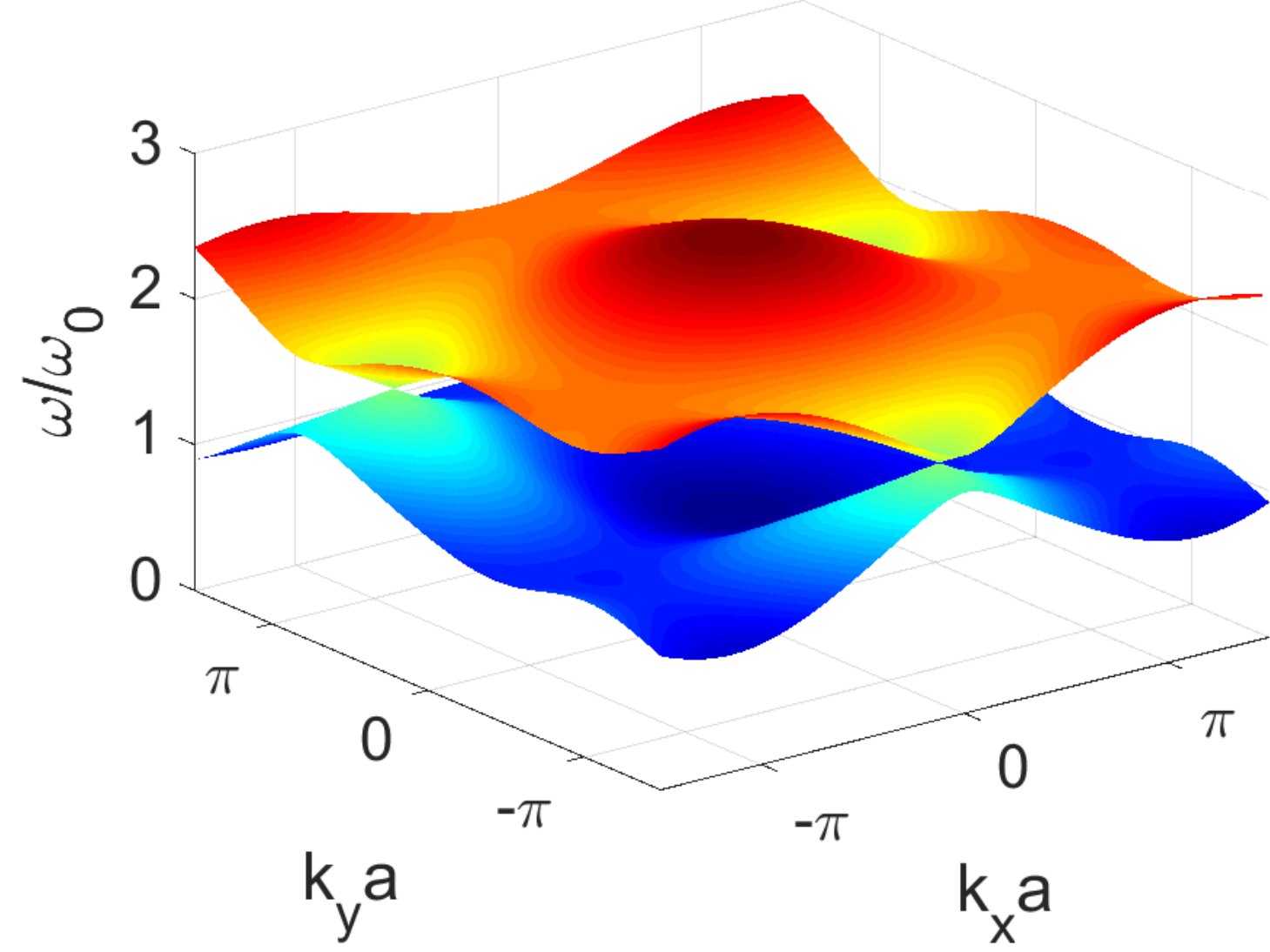} & \textbf{(b)} &
\includegraphics[width=5.4 cm, valign=t]{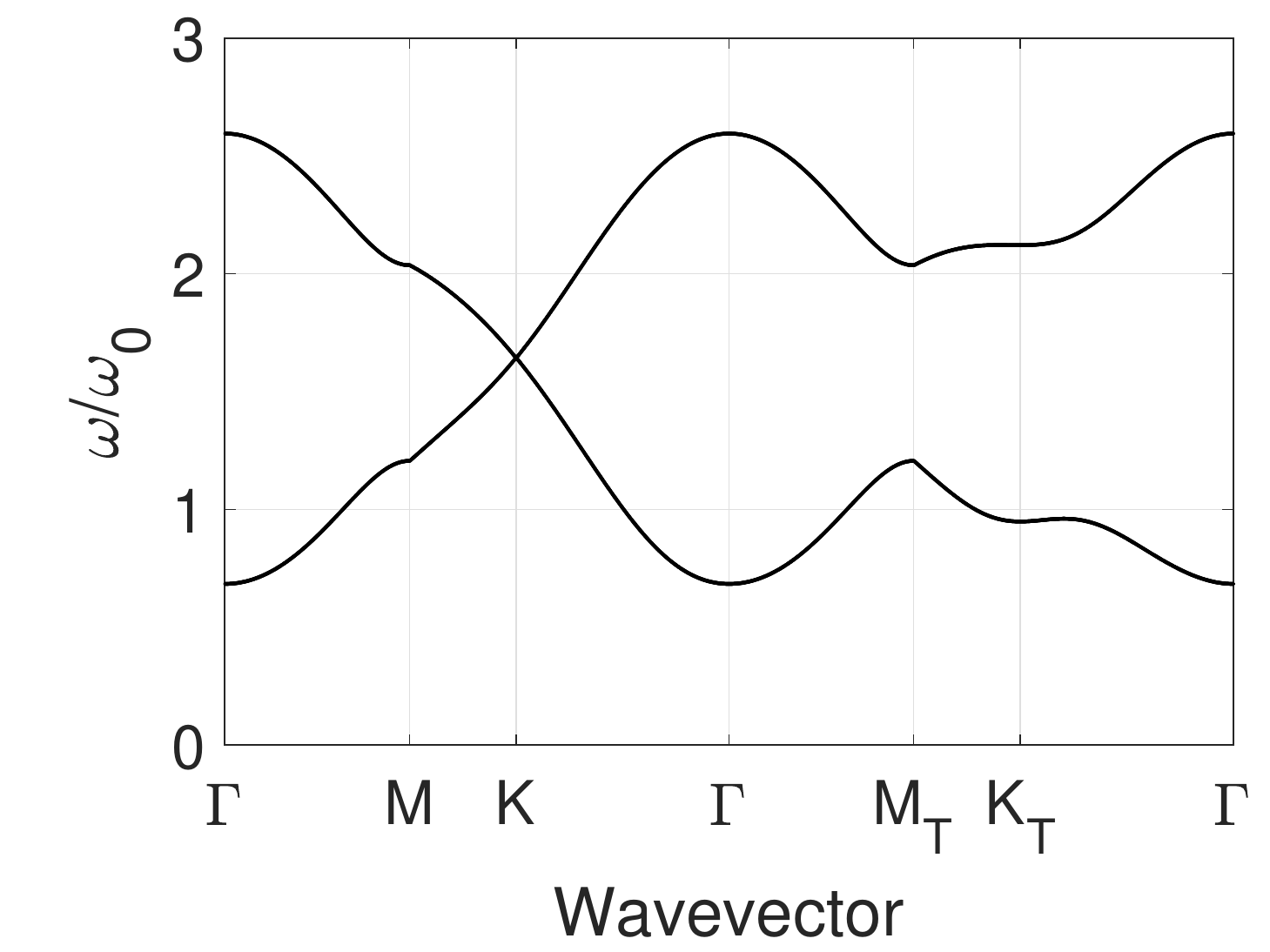} 
\end{tabular}
\caption{Infinite-lattice dispersion relations of the Haldane model mechanical analogue. Dispersion diagram with parameters $\{t_2=0.2t_1,\phi=\pi/3,M=3\sqrt{3}t_2\sin\phi\}$ is plotted over the entire Brillouin zone (a) and through the high symmetry points (b).}
\label{BulkDisp_param2}
\end{center}    
\end{figure}

\subsection{Modified Haldane model}

Here we provide derivation details for the mechanical analogue of the modified quantum Haldane model, which we realize with our feedback-based metamaterial in the main text. 
The transition from the original Haldane model to the modified one requires flipping the direction of the next-nearest-neighbor couplings in one of the unit cell sites, for example in the $B$ site, as illustrated in Fig. \ref{mHaldane_EdgeDisp}(a). This implies that for the $B$ site, the $t_2e^{-i\phi}$ couplings (blue arrows) are changed to $t_2e^{+i\phi}$ (red arrows), and vice versa. In order to create these new couplings in real-time, the corresponding closed-loop controller for the $B$ site takes the form
\begin{equation}    \label{eq:C_B}
C=\left( \begin{array}{ccccc} t_2\cos \phi & -\frac{t_2}{\omega}\sin \phi & 0 & 0 & 0 \\ 0 & 0 & t_2\cos \phi & \frac{t_2}{\omega}\sin \phi & 0 \\ 0 & 0 & 0 & 0 & - M \end{array} \right).
\end{equation}
The $A$ site controller remains unchanged. The dispersion profile of the resulting closed-loop semi-infinite system, a finite-sized strip periodic in the $x$ direction, is depicted in Fig. \ref{mHaldane_EdgeDisp}(b). 
Contrary to the original Haldane model, the bulk dispersion of the modified model is gapless. In addition, the two edge states have the same slope, which indicates that modes corresponding to these two edges have the same group velocity, and thus will propagate in the same direction, as we indeed demonstrate in the dynamical simulations in Fig. 4(c),(d) in the main text. 

\begin{figure}[htpb] 
\begin{center}
\begin{tabular}{l l l l}
   \textbf{(a)} & \includegraphics[width=6.6 cm, valign=t]{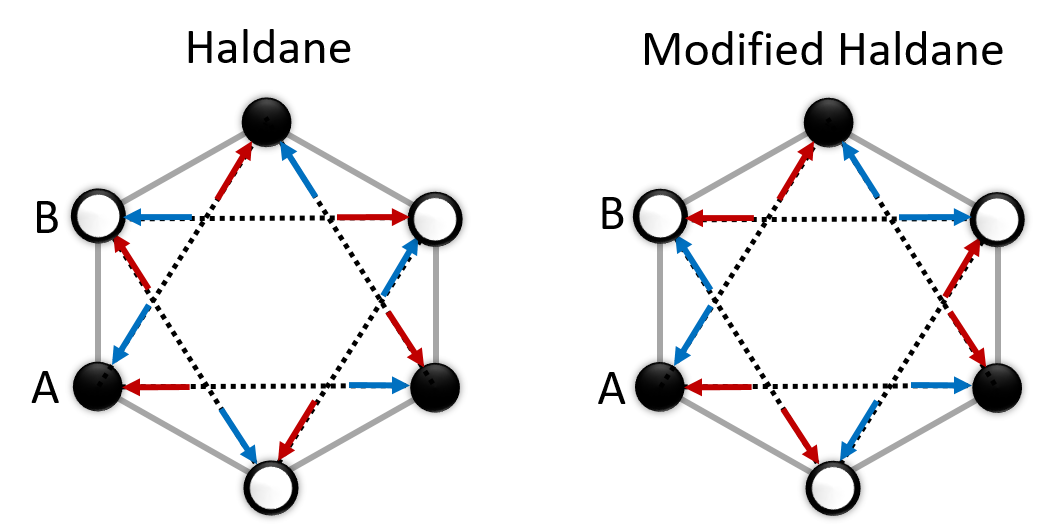} & \textbf{(b)}  &  
\includegraphics[width=6.9 cm, valign=t]{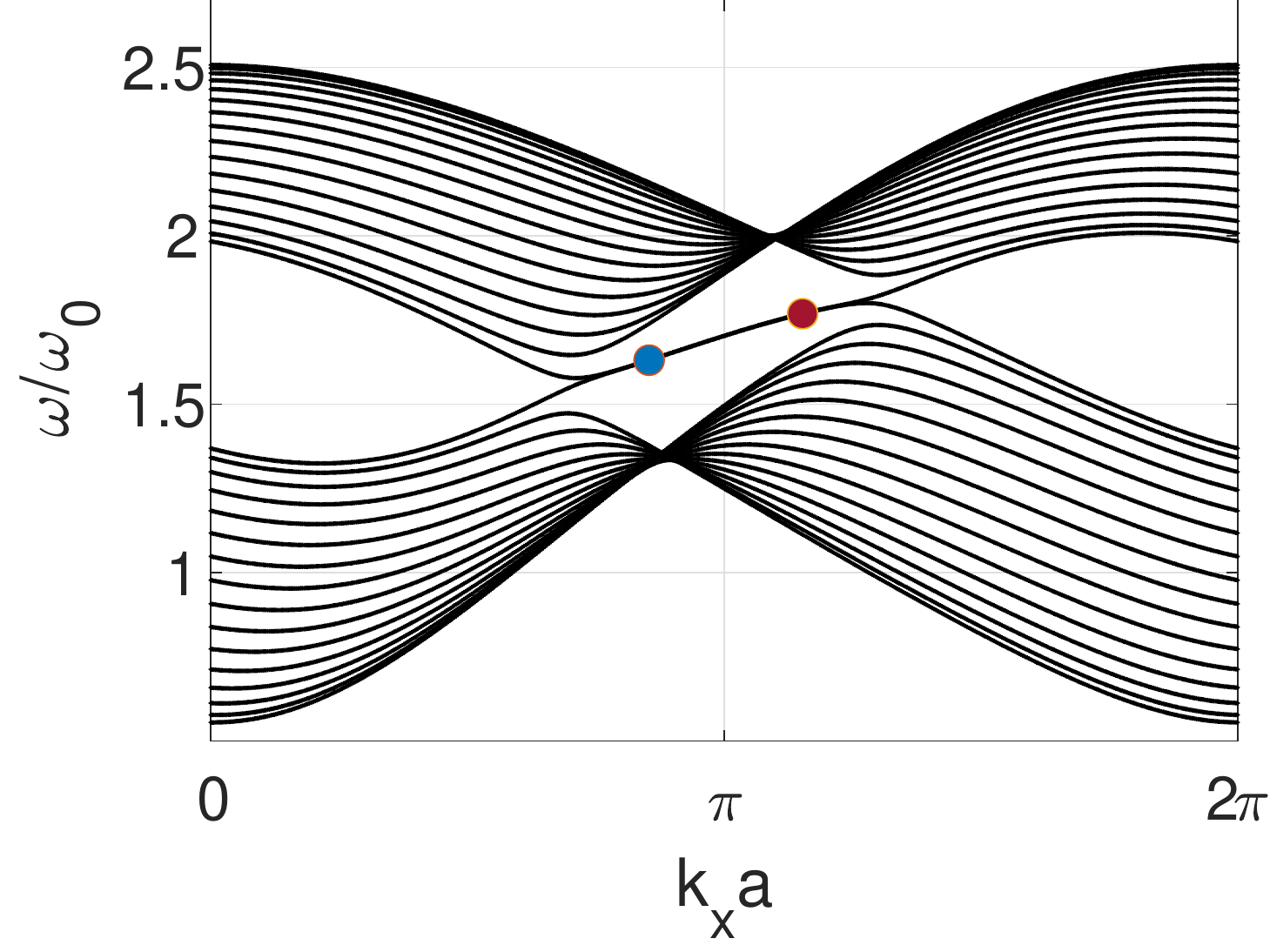} 
\end{tabular}
\caption{The modified Haldane model. Unit cell schematic (modified vs. original) (a), and semi-infinite lattice dispersion profile (b).}
\label{mHaldane_EdgeDisp}
\end{center}    
\end{figure}

\subsection{Pseudospin multipole topological insulator}

In this section we demonstrate the versatility of the feedback-based design method presented in the main text, and its ability to use the same physical platform to implement different quantum topological wave phenomena. In particular, we use feedback control to realize a mechanical analogy of the quantum spin Hall effect (QSHE), i.e. a topological insulator, on the same host structure that was used in the main text to realize the Haldane model quantum Hall effect (a Haldane Chern insulator), and the modified Haldane model. Switching between these realizations is carried out exclusively through the control program. \\
To this end we derive a closed-loop control strategy to create inter-site couplings, which turns the metamaterial presented in the main text into a topological insulator emulating the QSHE.
The underlying pseudospin-orbit coupling can be obtained via spatial symmetry breaking in the lattice structure.
In acoustic or mechanical metamaterials this was realized by various configurations [19-21, 27,31], where actual spinning DOFs, such as in-plane rotating masses or circulating acoustic velocity fields were used (or an equivalent two DOFs per site setting [22]). 
Here we realize the pseudospin modes on a single DOF per site platform, i.e. an out-of-plane vibration with no physical rotation. \\
We convert the bipartite honeycomb of lattice constant $a'$ into a six-site unit cell honeycomb of lattice constant $a$, outlined by the light blue hexagons in Fig. \ref{QSHE_scheme}(a).
This requires the intra-cell couplings of the expanded cells to differ from the inter-cell ones. 
We therefore program the embedded control system to create couplings according to the red arrows in Fig. \ref{QSHE_scheme}(a), by using measurements of the end masses relative displacements. The closed loop operation for the expanded unit cell is illustrated in Fig. \ref{QSHE_scheme}(b). We program the controller block to create real-valued intra-cell couplings of strength $t_2\neq t_1$, instead of the $t_1$ strength in the actual lattice. The new $t_2$ couplings, which are equivalent to linear springs of stiffness $t_2$, are created via the control action alone, and are attained in real-time.
The governing equations of the open loop uniform lattice, the control law for the forces, and the control matrix $C$ for the $\{i,j\}$ unit cell, are then given by
\begin{equation}   \label{eq:Motion_OL}
\begin{cases}
\ddot{u}^1_{i,j}=t_1\left(u^2_{i,j}+u^4_{i,j+1}+u^6_{i,j}\right)-3t_1u^1_{i,j}+f^1_{i,j} \\
\ddot{u}^2_{i,j}=t_1\left(u^1_{i,j}+u^3_{i,j}+u^5_{i+1,j}\right)-3t_1u^2_{i,j}+f^2_{i,j} \\
\ddot{u}^3_{i,j}=t_1\left(u^2_{i,j}+u^4_{i,j}+u^6_{i+1,j-1}\right)-3t_1u^3_{i,j}+f^3_{i,j} \\
\ddot{u}^4_{i,j}=t_1\left(u^1_{i,j-1}+u^3_{i,j}+u^5_{i,j}\right)-3t_1u^4_{i,j}+f^4_{i,j} \\
\ddot{u}^5_{i,j}=t_1\left(u^2_{i-1,j}+u^4_{i,j}+u^6_{i,j}\right)-3t_1u^5_{i,j}+f^5_{i,j} \\
\ddot{u}^6_{i,j}=t_1\left(u^1_{i,j}+u^3_{i-1,j+1}+u^5_{i,j}\right)-3t_1u^6_{i,j}+f^6_{i,j}
\end{cases}, \qquad \left(\begin{array}{l}
f^1_{i,j} \\ f^2_{i,j} \\f^3_{i,j} \\f^4_{i,j} \\ f^5_{i,j} \\ f^6_{i,j}
\end{array}\right)=C\left(\begin{array}{l}
u^4_{i,j+1}-u^1_{i,j} \\ u^5_{i+1,j}-u^2_{i,j} \\ u^6_{i+1,j-1}-u^3_{i,j} \\ u^1_{i,j-1}-u^4_{i,j} \\ u^2_{i-1,j}-u^5_{i,j} \\ u^3_{i-1,j+1}-u^6_{i,j}
\end{array}\right), \qquad C=\left(t_2-t_1\right)I_6,
\end{equation}
where $I_6$ is a $6x6$ identity matrix.
The expansion of the metamaterial unit cell in real space results in a corresponding folding of the wavevector $\textbf{k}$ space, as illustrated in Fig. \ref{QSHE_scheme}(c), with folding lines given in dashed blue. The closed (open) loop Brillouin zone is portrayed in solid blue (black), with the high symmetry points given by $\Gamma$ ($\Gamma'$), $K$ ($K'$) and $M$ ($M'$). 
The blue Brillouin zone is used to calculate the frequency dispersion relation $\omega(\textbf{k})$ of an infinite-sized closed loop metamaterial.
The corresponding eigenvalue problem with the eigenstate $\textbf{q}=[\begin{array}{c c c c c c}
q^1 & q^2 & q^3 & q^4 & q^5 & q^6 \end{array}]^T$ and the effective Hamiltonian $H(\textbf{k})$, are given by
\begin{figure}[b] 
\begin{center}
\begin{tabular}{l l l}
\small{\textbf{(a)}} & \small{\textbf{(b)}} & \small{\textbf{(c)}} \\
\includegraphics[height=4.7 cm, valign=c]{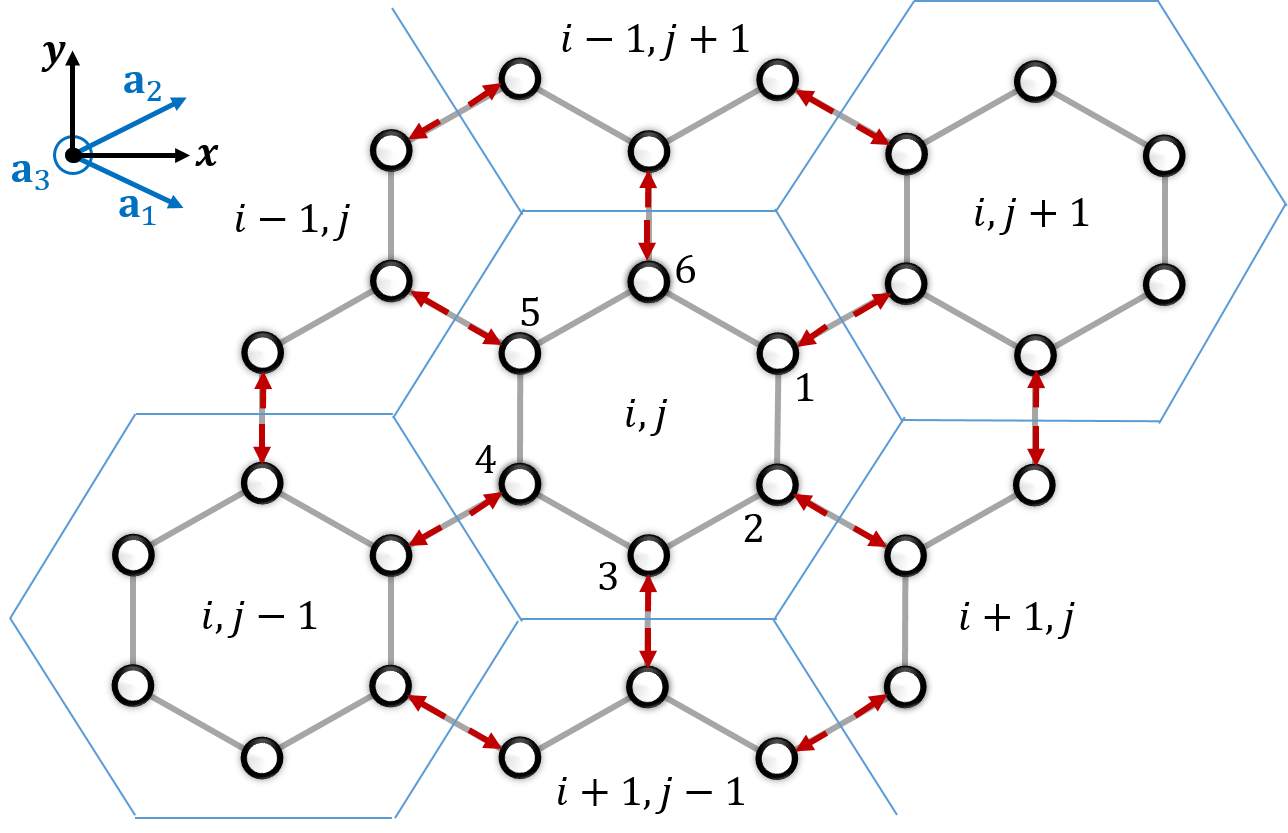} & \includegraphics[height=3.8 cm, valign=c]{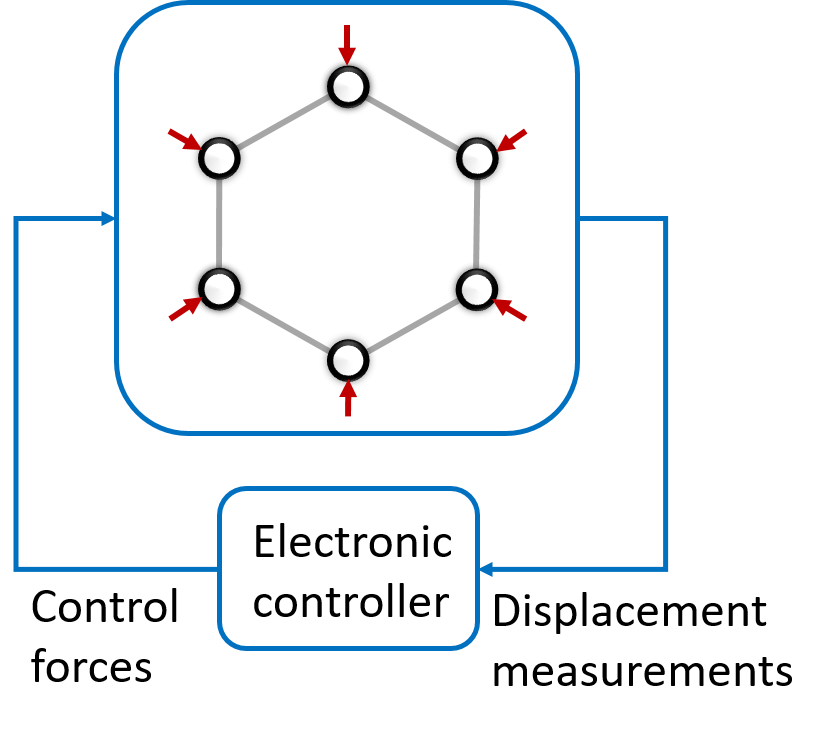} & \includegraphics[height=3.1 cm, valign=c]{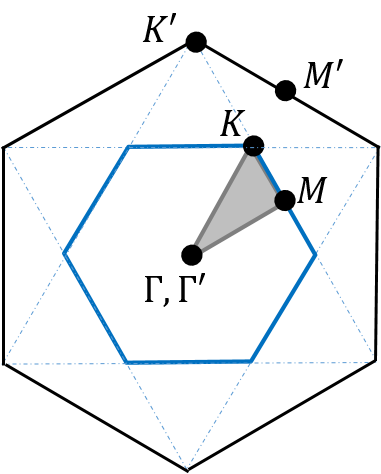} 
\end{tabular}
\caption{(a) Schematic of the mechanical metamaterial of the main text, which is designed here to mimic the QSHE using our feedback-based method. Circles and grey bars are identical masses and linear springs, respectively. Red arrows indicate the measurements required for the control loops. (b) The feedback-based mechanism for the corresponding six-site unit cell. (c) The resulting folded Brillouin zone.}
\label{QSHE_scheme}
\end{center}
\end{figure}
\begin{equation}  \label{eq:EQ}
\omega^2\textbf{q}(\textbf{k})=H(\textbf{k})\textbf{q}(\textbf{k}), \qquad H(\textbf{k})=\left(\begin{array}{c c c c c c}
2t_1+t_2 & -t_1 & 0 & -t_2e^{i\textbf{k}\cdot\textbf{a}_2} & 0 & -t_1 \\
-t_1 & 2t_1+t_2 & -t_1 & 0 & -t_2e^{i\textbf{k}\cdot\textbf{a}_1} & 0 \\
0 & -t_1 & 2t_1+t_2 & -t_1 & 0 & -t_2e^{i\textbf{k}\cdot\left(\textbf{a}_1-\textbf{a}_2\right)} \\
-t_2e^{-i\textbf{k}\cdot\textbf{a}_2} & 0 & -t_1 & 2t_1+t_2 & -t_1 & 0 \\
0 & -t_2e^{-i\textbf{k}\cdot\textbf{a}_1} & 0 & -t_1 & 2t_1+t_2 & -t_1 \\
-t_1 & 0 & -t_2e^{-i\textbf{k}\cdot\left(\textbf{a}_1-\textbf{a}_2\right)} & 0 & -t_1 & 2t_1+t_2
\end{array}\right).
\end{equation}

We now demonstrate that the metamaterial governed by \eqref{eq:EQ} can exhibit a topological insulator. 
The key principle includes the relative values of $t_1$ and $t_2$. Artificially regarding these spring constants with respect to the distances between the corresponding end masses, results in a geometrical constraint, which implies
\begin{equation}  \label{eq:t1t2}
t_1=\frac{t_0}{3\beta a_0}, \qquad \qquad t_2=\frac{t_0}{3\left(1-2\beta\right)a_0},
\end{equation}
where $t_0$ is a nominal spring constant normalized by $m_0$, $a_0=\frac{1}{3}a$ is a nominal distance and $\beta\in\left(0,\frac{1}{2}\right)$ is a design parameter. We then obtain $t_1<t_2$ for $\beta<\frac{1}{3}$, and $t_1>t_2$ for $\beta>\frac{1}{3}$. Choosing the values $a_0=0.05\sqrt{3}$ and $t_0\approx 1.4\cdot 10^5$, we solve the eigenvalue problem in \eqref{eq:EQ}. 
The resulting frequency dispersion of the infinite closed loop metamaterial is depicted in Fig. \ref{QSHE_infinite}(a),(b) and (c) for $\beta=\frac{1}{3}$, $\beta=0.7\cdot\frac{1}{3}$ and $\beta=1.2\cdot\frac{1}{3}$, respectively. 
For $\beta=\frac{1}{3}$, i.e. for $t_1=t_2$, we obtain a gapless spectrum, which is the folded spectrum of a two-site honeycomb lattice. A double Dirac-like cone appears at the $\Gamma$ point, indicated by a black dot in Fig. \ref{QSHE_infinite}(a). For both $\beta>\frac{1}{3}$ ($t_1>t_2$) and $\beta<\frac{1}{3}$ ($t_1<t_2$) a gap is opened at the $\Gamma$ point, bounded by two pairs of bands that are degenerate at the $d$ (black circle) and $p$ (black square) frequencies. Due to the geometrical rule in \eqref{eq:t1t2} the two gaps are overlapping.
\begin{figure}[h] 
\begin{center}
\begin{tabular}{l l l}
\small{\textbf{(a)} $t_1=t_2$} &  \small{\textbf{(b)} $t_1>t_2$, trivial (T)} & \small{\textbf{(c)}  $t_1<t_2$, non-trivial (NT)} \\
\includegraphics[height=4.1 cm, valign=c]{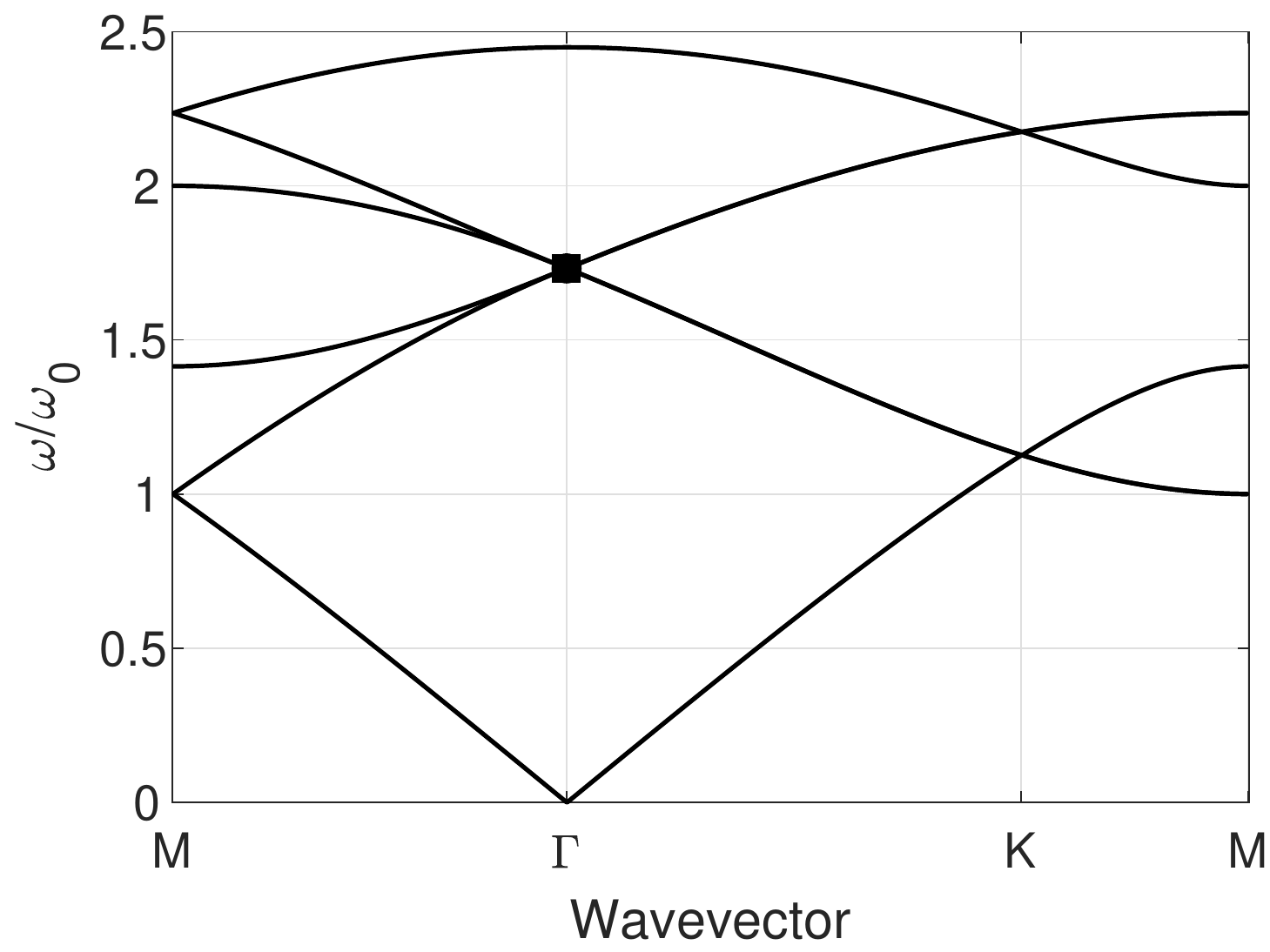} & \includegraphics[height=4.1 cm, valign=c]{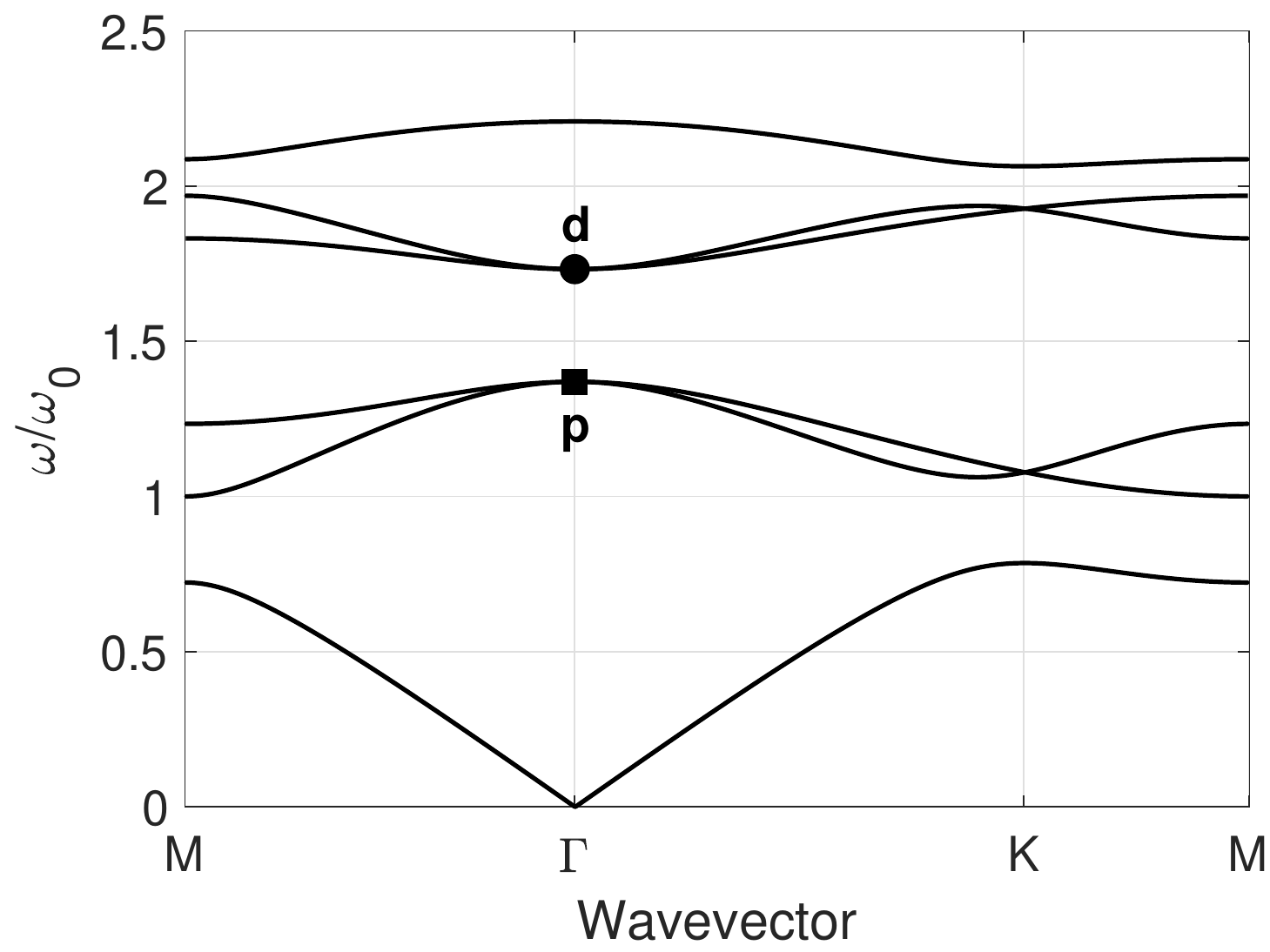} &  \includegraphics[height=4.1 cm, valign=c]{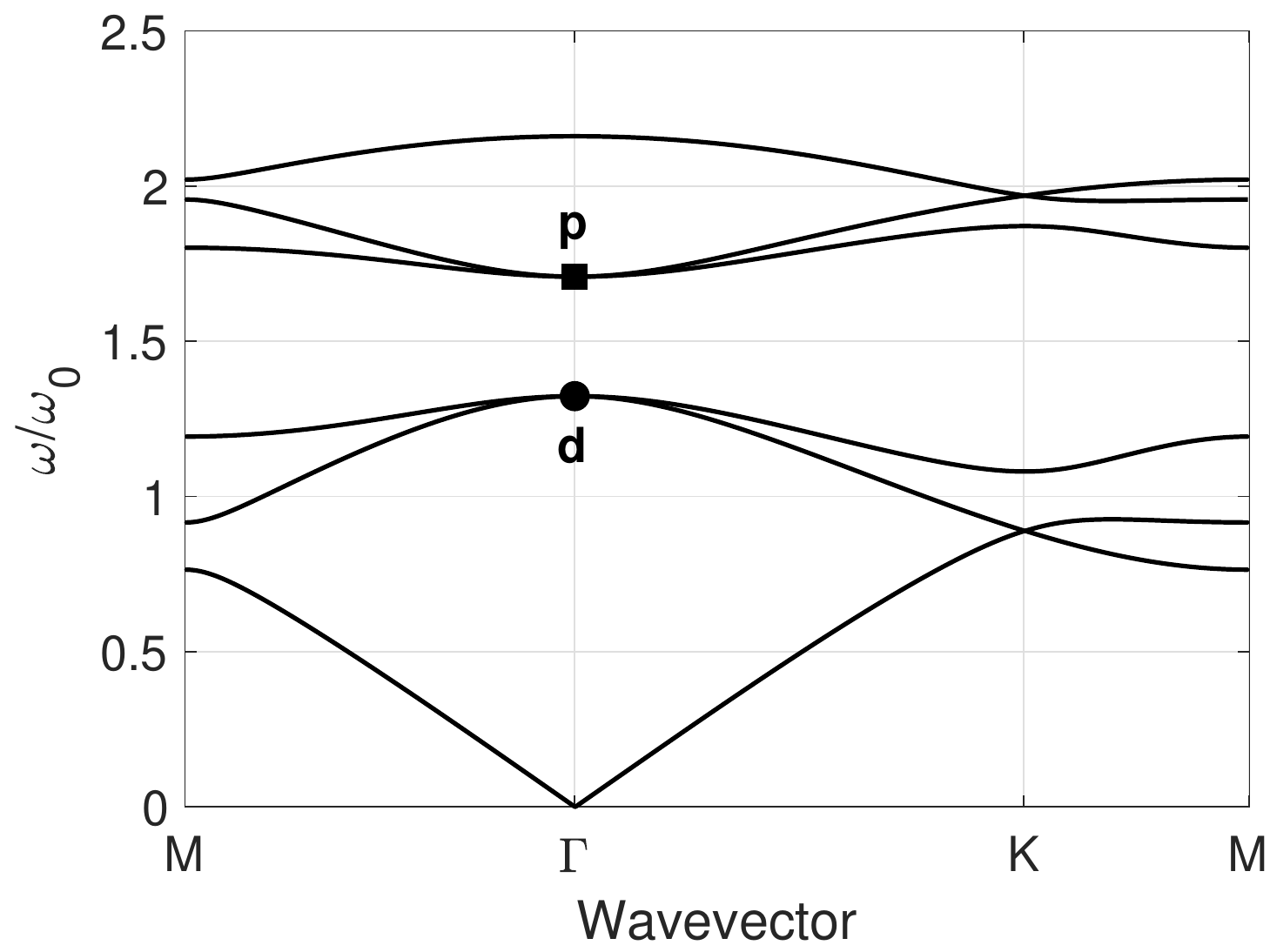}
\end{tabular}
\caption{Dispersion profile of the infinite metamaterial. (a) The nominal system with all identical spring constants $t_1=t_2$, $\beta=\frac{1}{3}$. A double Dirac-like cone emerges at the $\Gamma$ point. (b),(c) The controlled metamaterial in the topologically trivial (non-trivial) regime with $t_1>t_2$, $\beta>\frac{1}{3}$ ($t_1<t_2$, $\beta<\frac{1}{3}$). A band-gap emerges at the $\Gamma$ point. The dipole modes $p_x$, $p_y$, indicated by a black square, are below (above) the quadrupole modes $d_{xy}$, $d_{x^2-y^2}$, indicated by a black circle. The transition from the trivial to the non-trivial topological regime involves band inversion.}
\label{QSHE_infinite}
\end{center}
\end{figure}

\begin{figure}[t] 
\begin{center}
\begin{tabular}{l l l l}
\setlength{\tabcolsep}{0pt}
\small{\textbf{(a)}} \includegraphics[height=0.6 cm, valign=t]{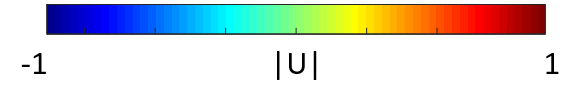} &  \small{\textbf{(b)}} \includegraphics[height=0.3 cm, valign=t]{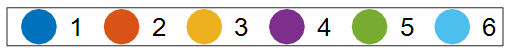} & \small{\textbf{(c)}} \includegraphics[height=0.6 cm, valign=t]{m_legend} & \small{\textbf{(d)}} \includegraphics[height=0.3 cm, valign=t]{n_legend}\\
\includegraphics[height=3.0 cm, valign=t]{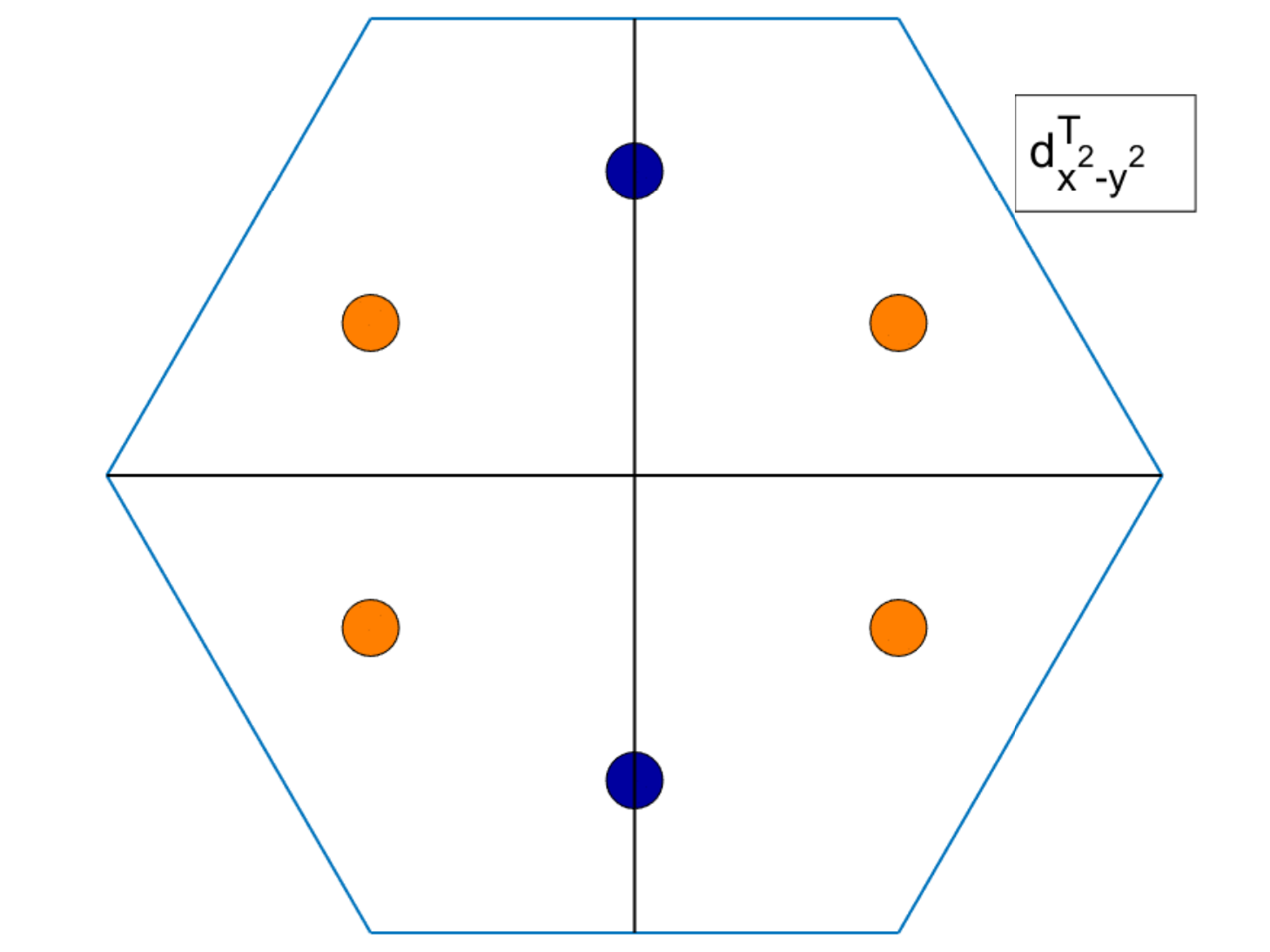} & \includegraphics[height=3.2 cm, valign=t]{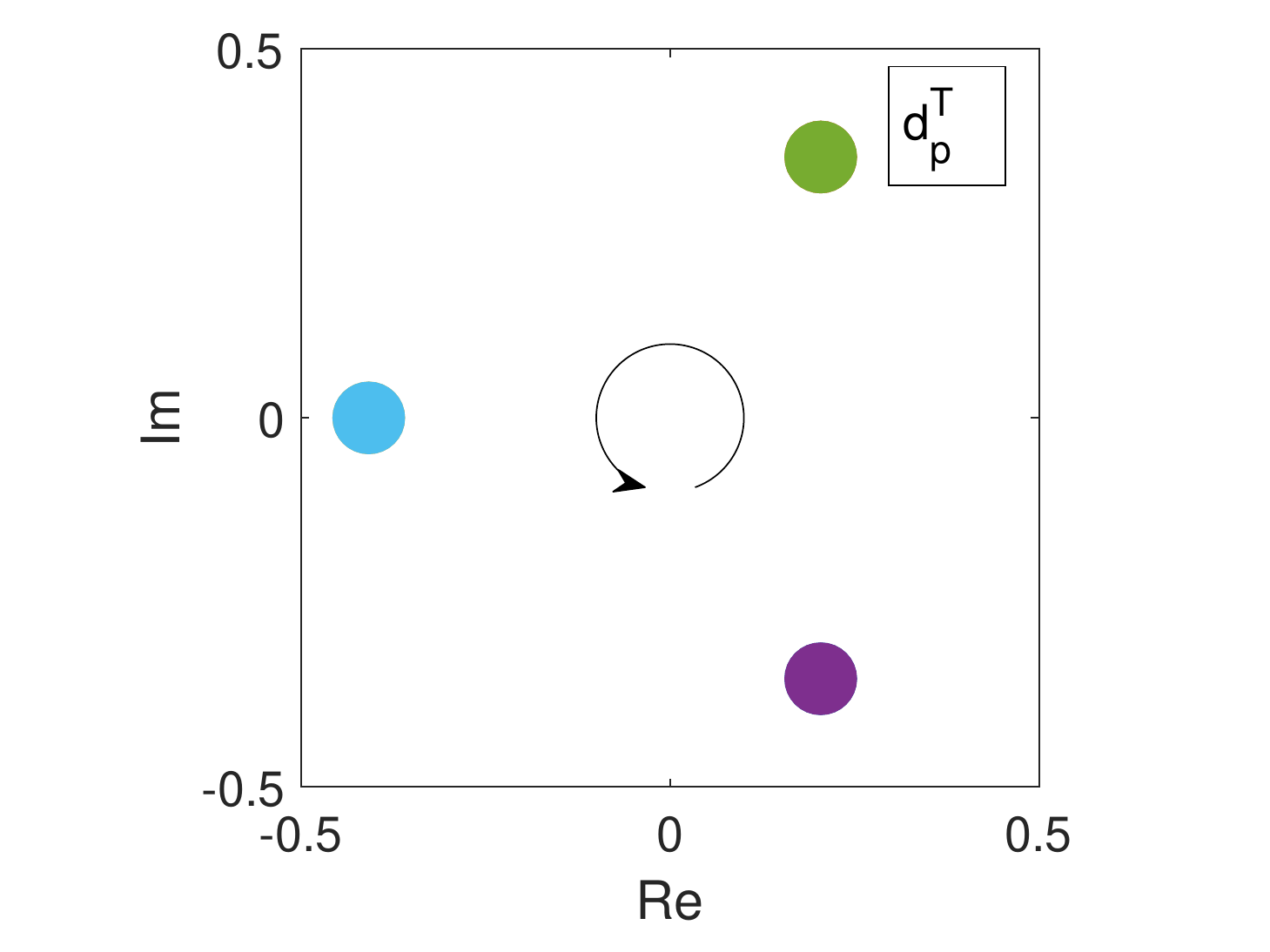} &  \includegraphics[height=3.0 cm, valign=t]{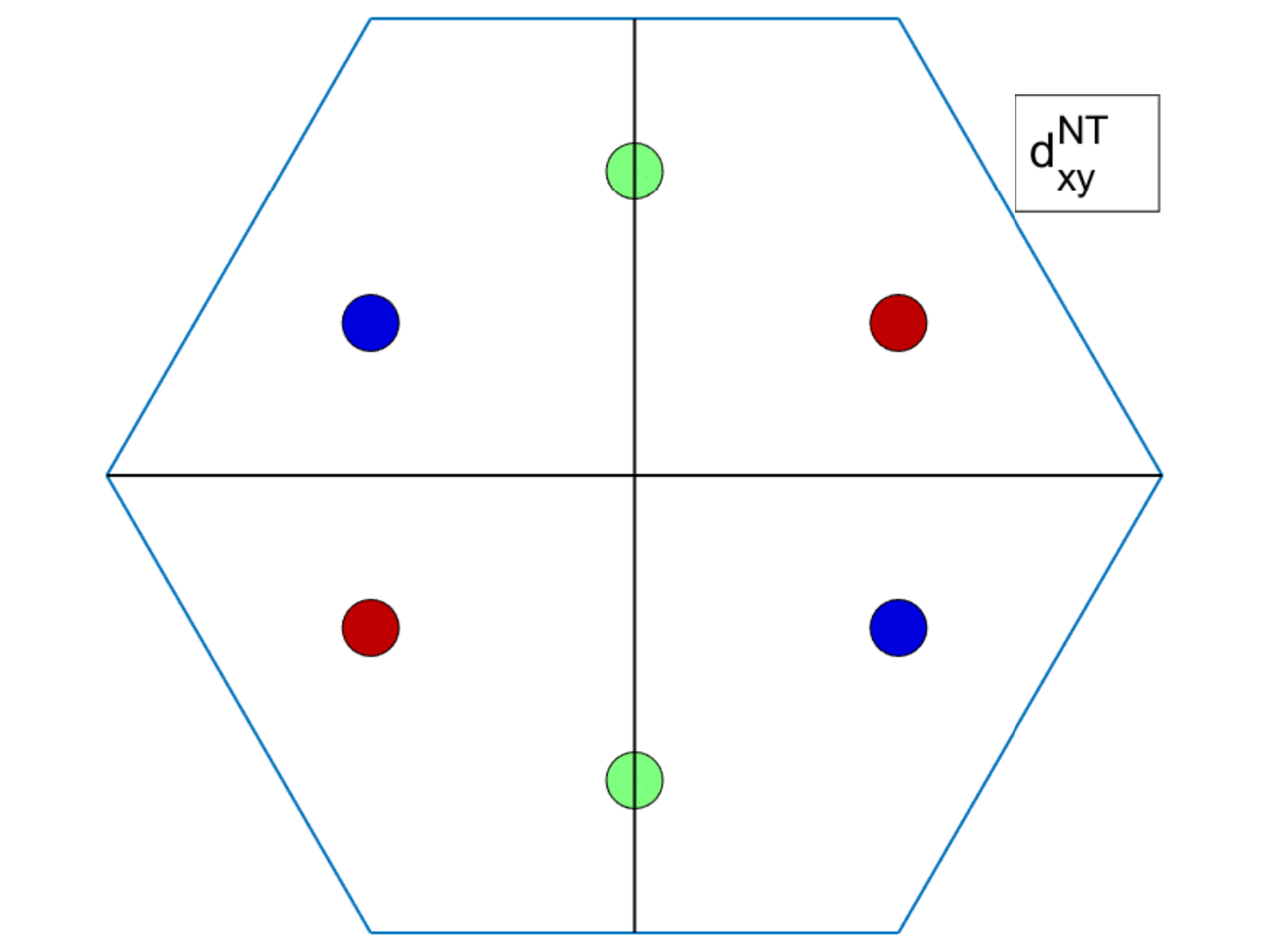} & \includegraphics[height=3.2 cm, valign=t]{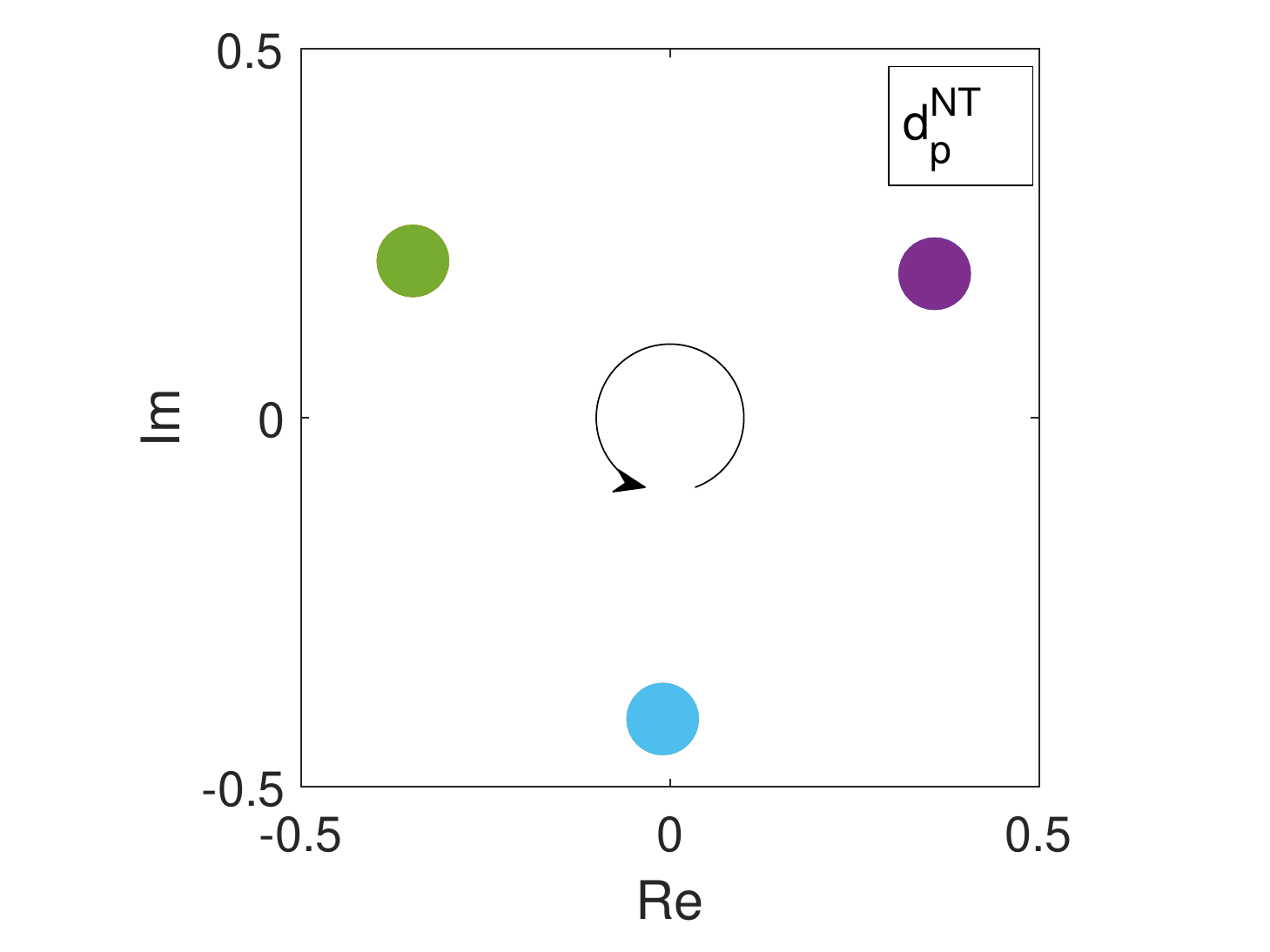} \\
\includegraphics[height=3.0 cm, valign=t]{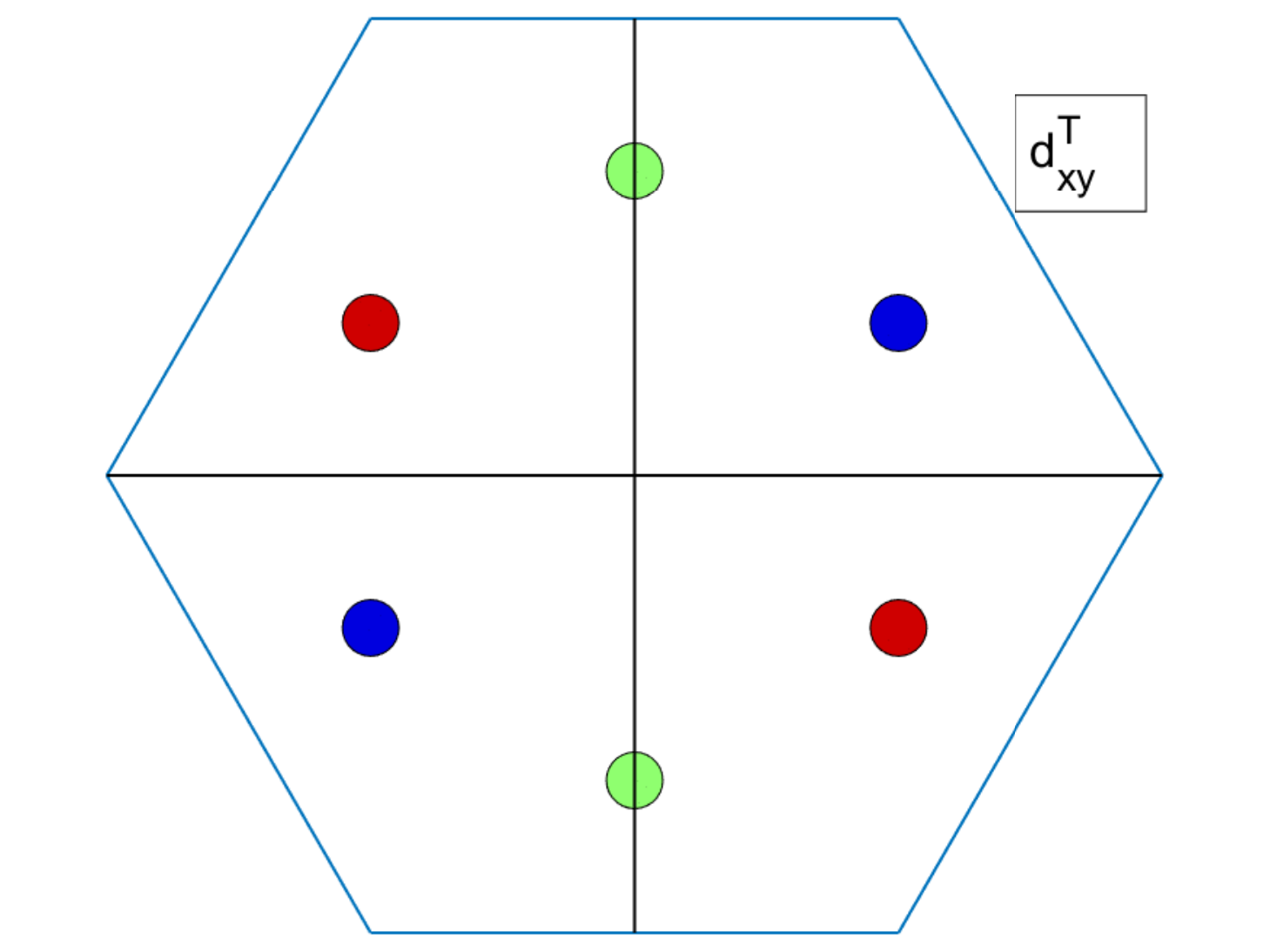} & \includegraphics[height=3.2 cm, valign=t]{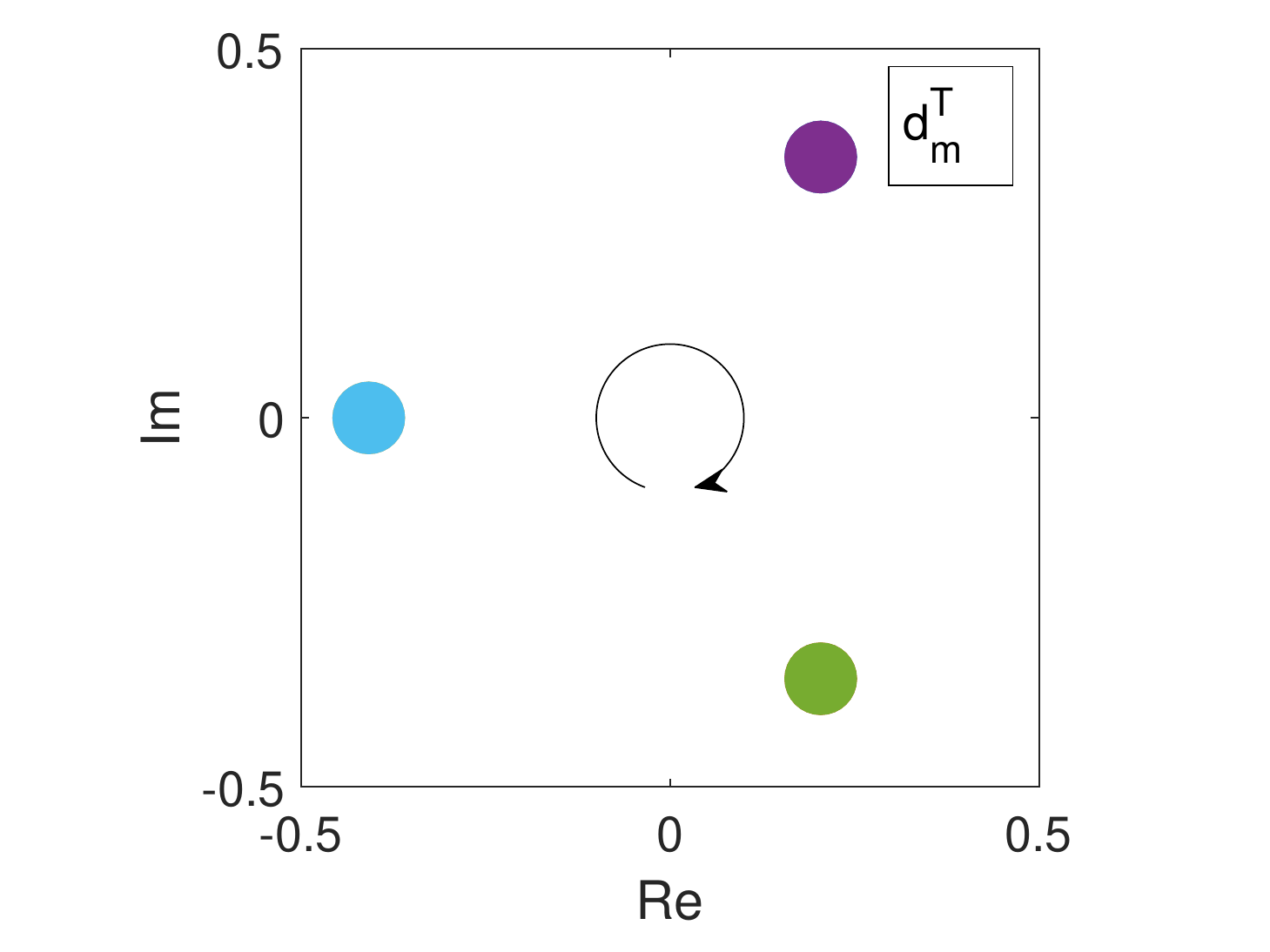} &  \includegraphics[height=3.0 cm, valign=t]{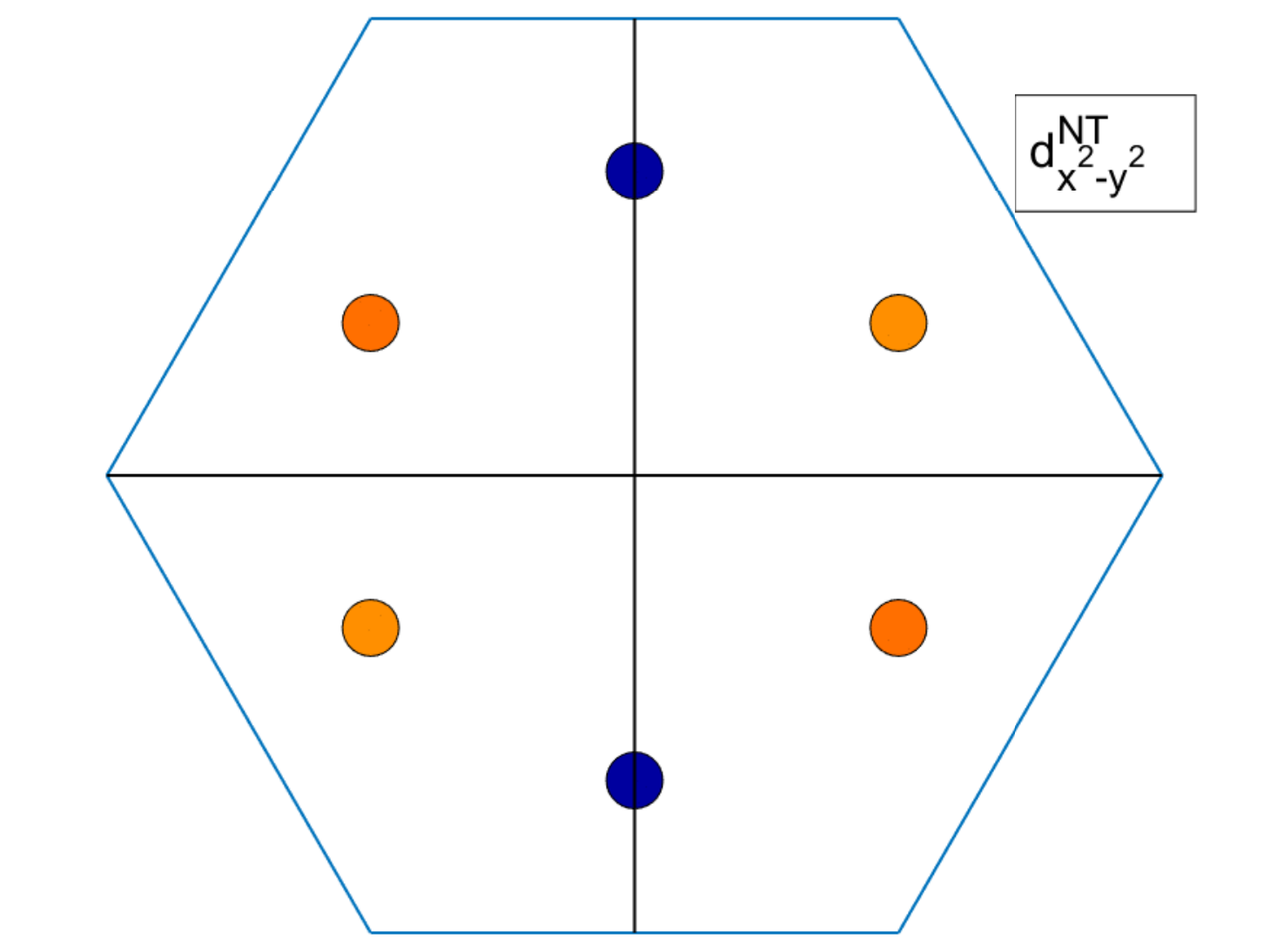} & \includegraphics[height=3.2 cm, valign=t]{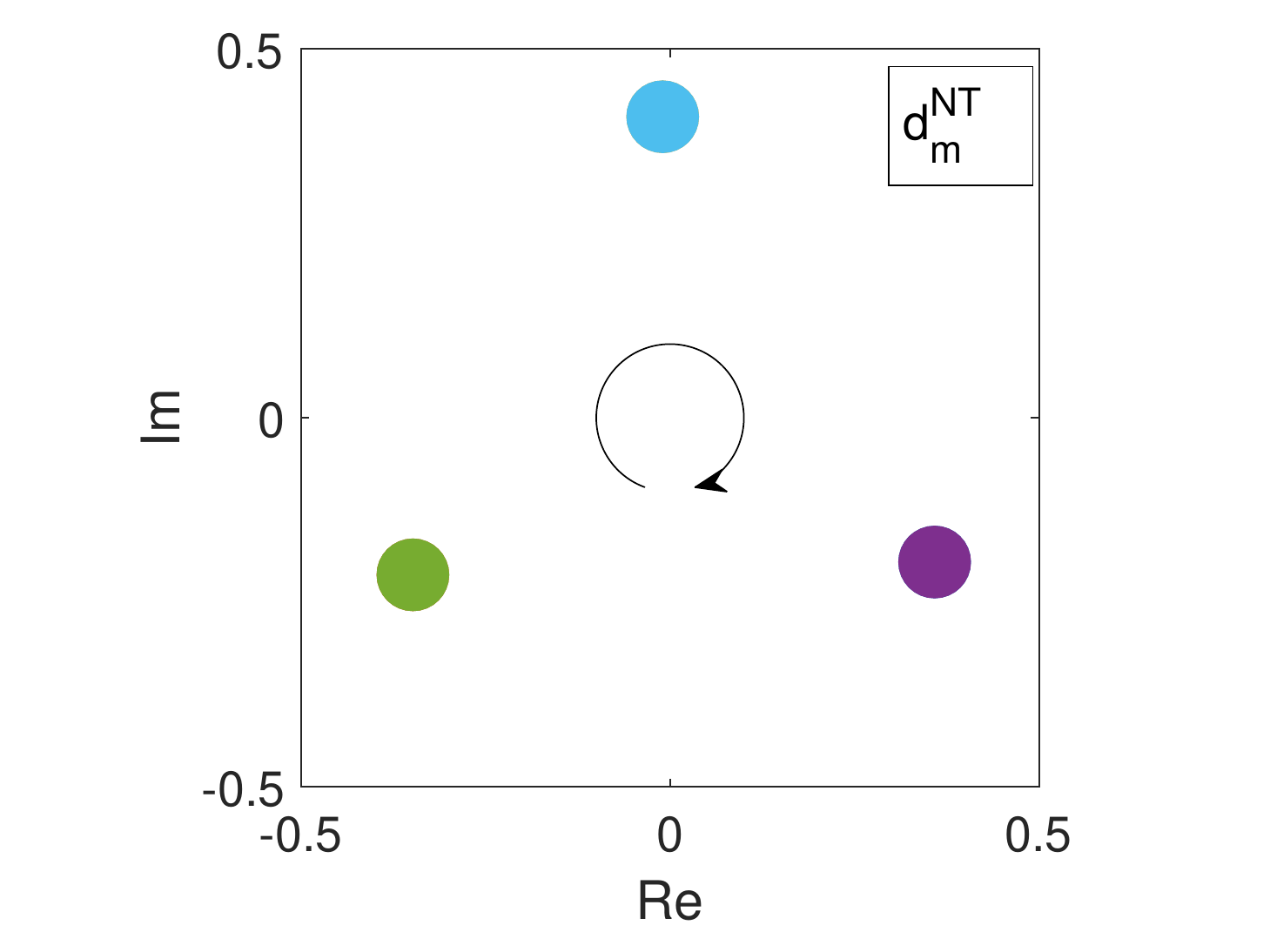} \\
\includegraphics[height=3.0 cm, valign=t]{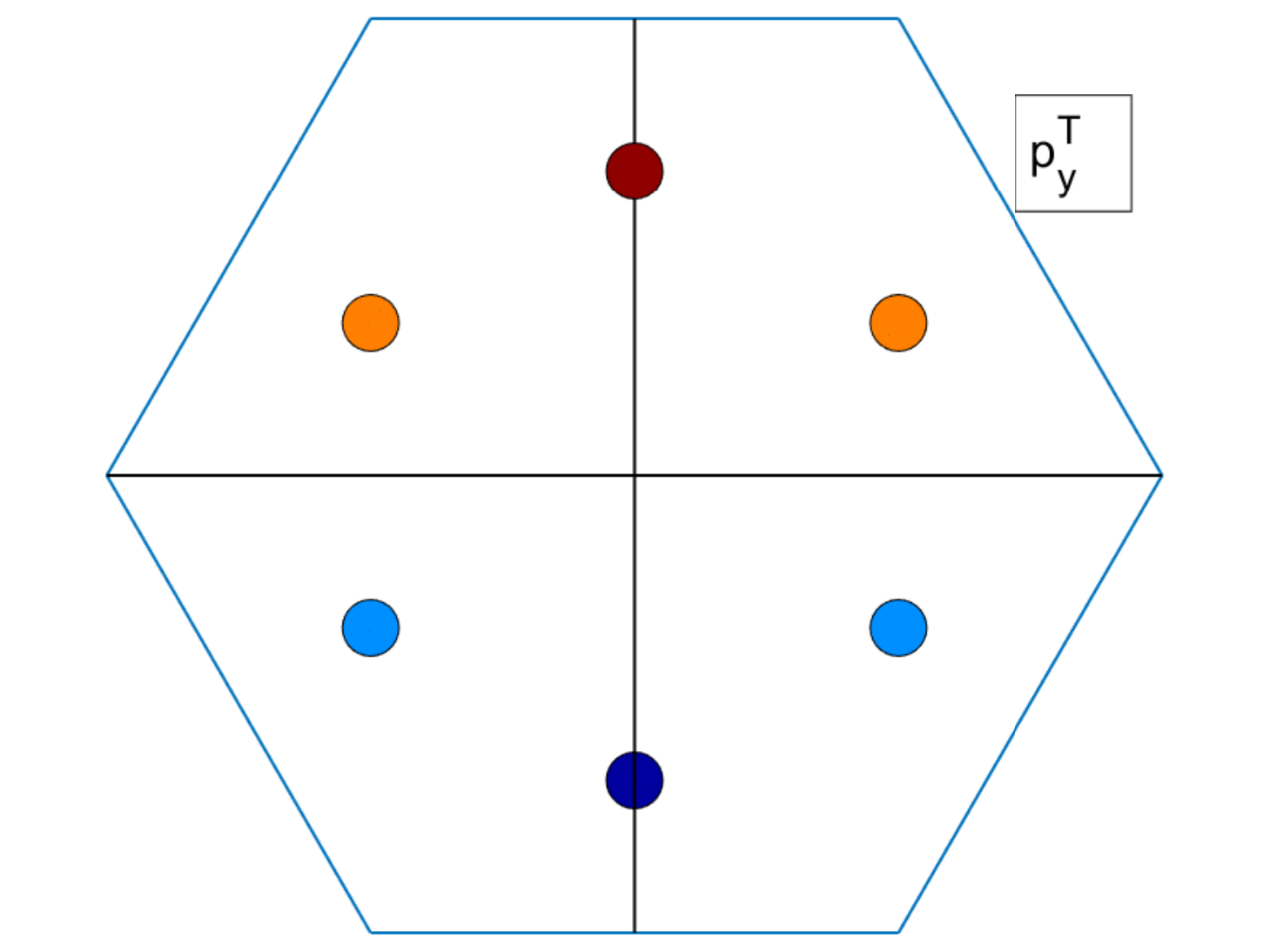} & \includegraphics[height=3.2 cm, valign=t]{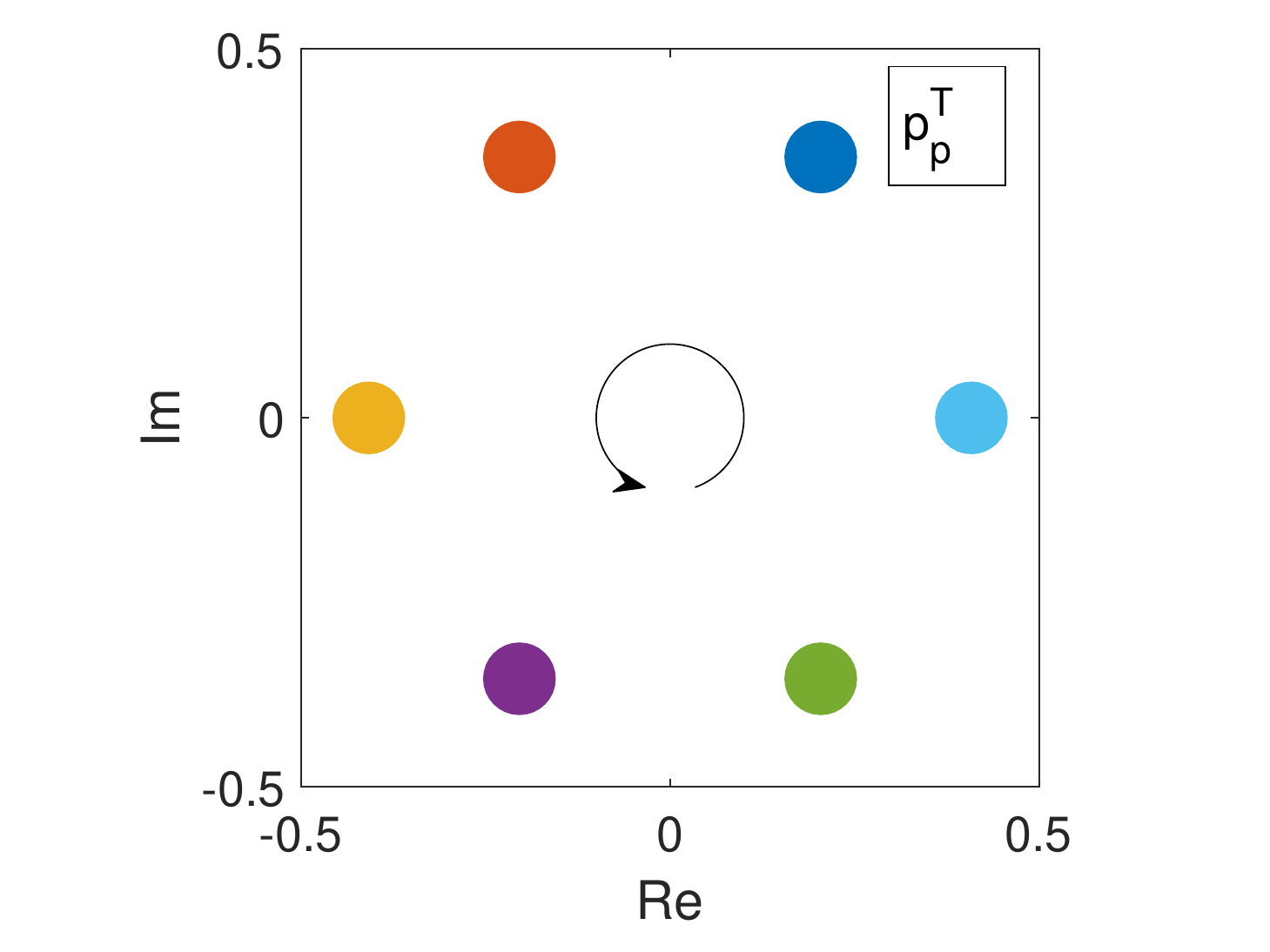} &  \includegraphics[height=3.0 cm, valign=t]{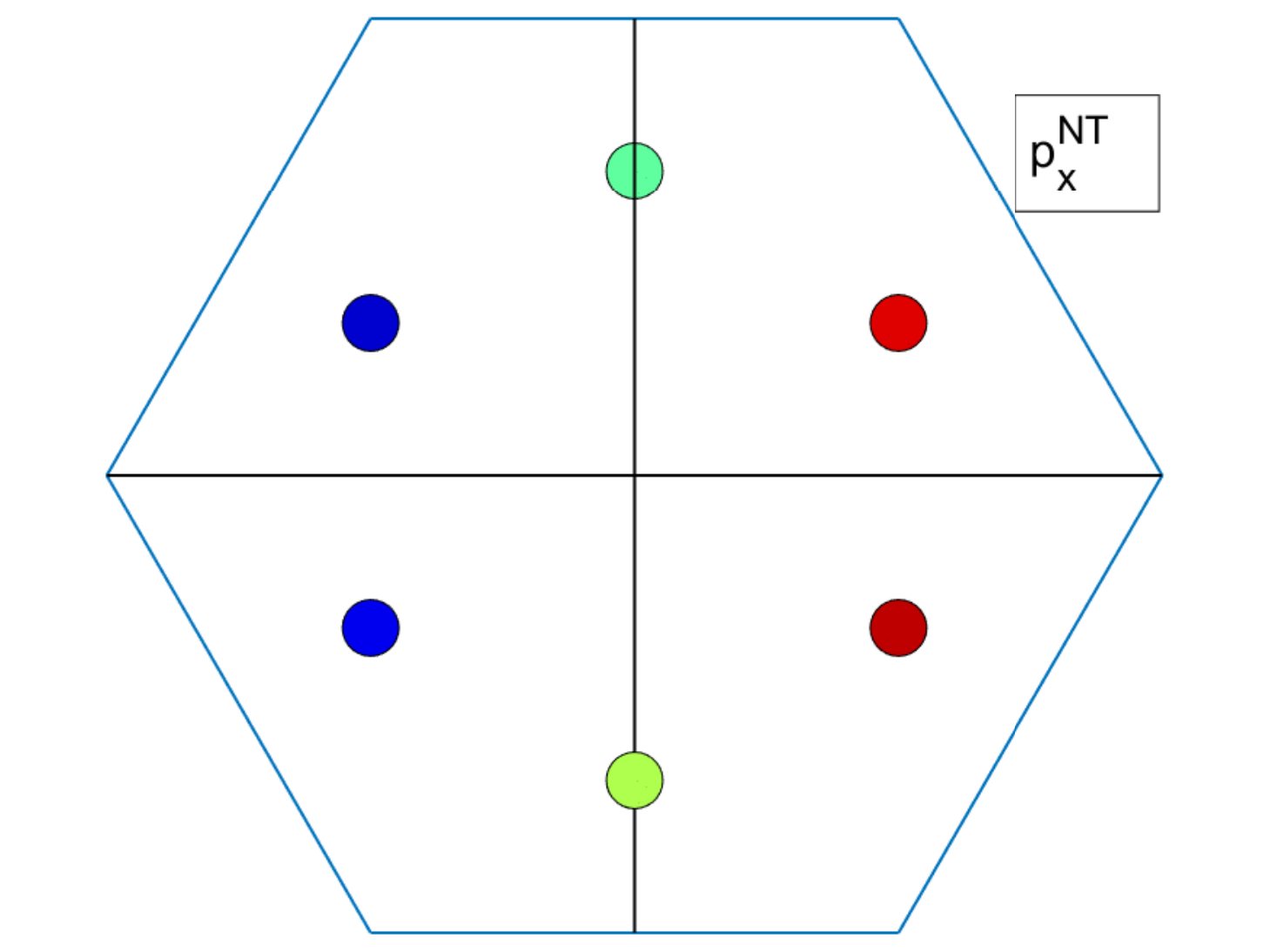} & \includegraphics[height=3.2 cm, valign=t]{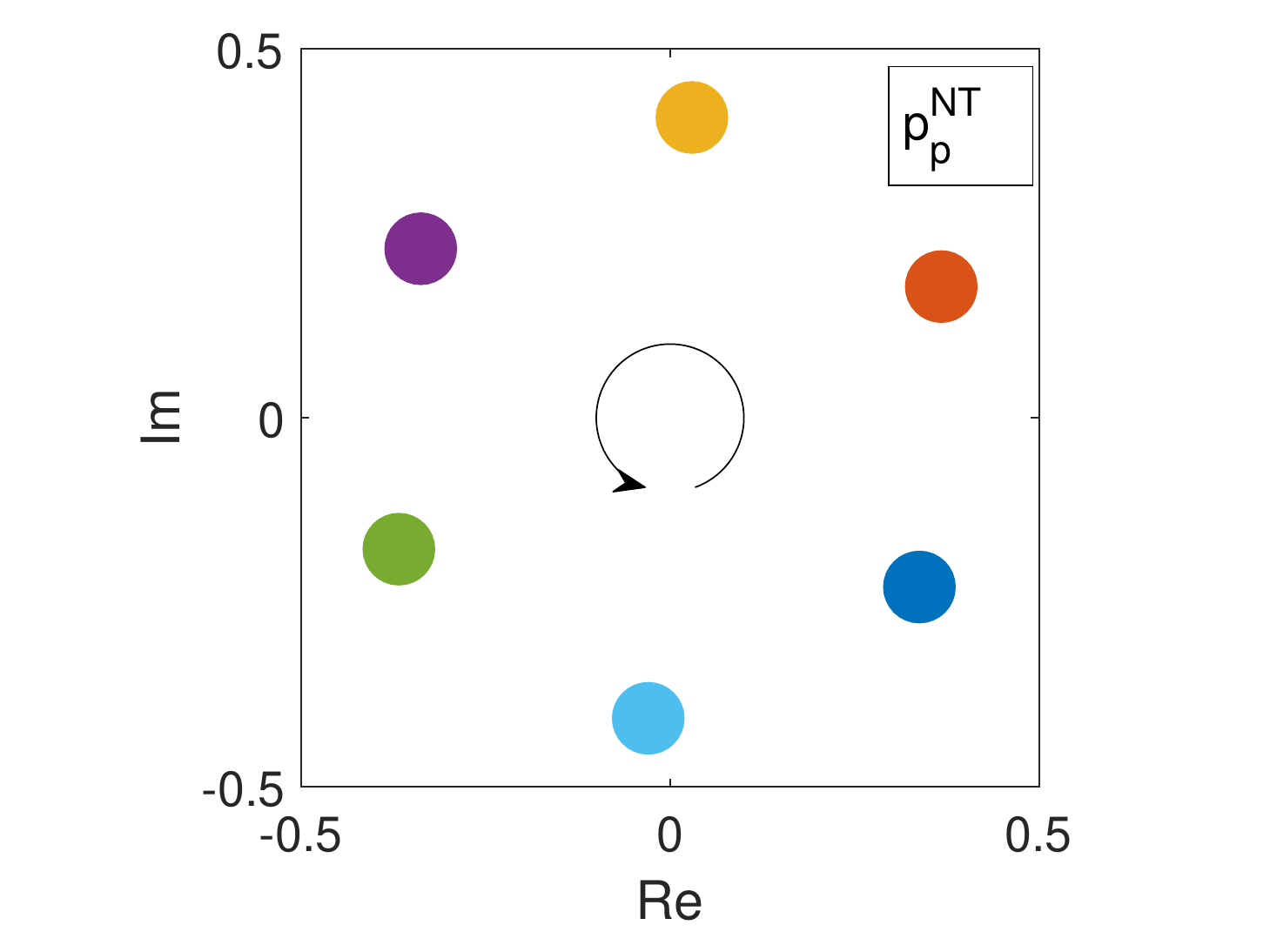} \\
\includegraphics[height=3.0 cm, valign=t]{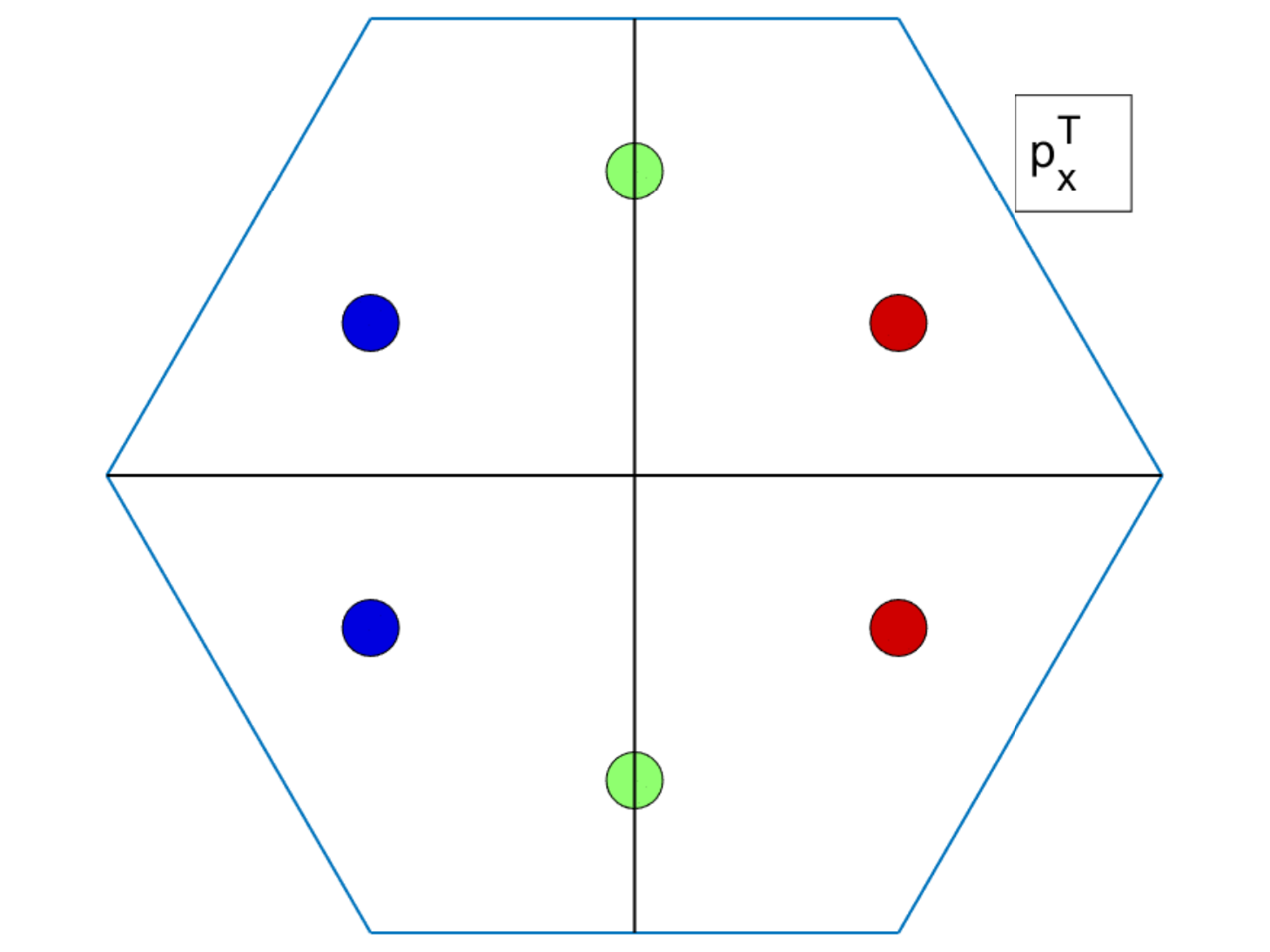} & \includegraphics[height=3.2 cm, valign=t]{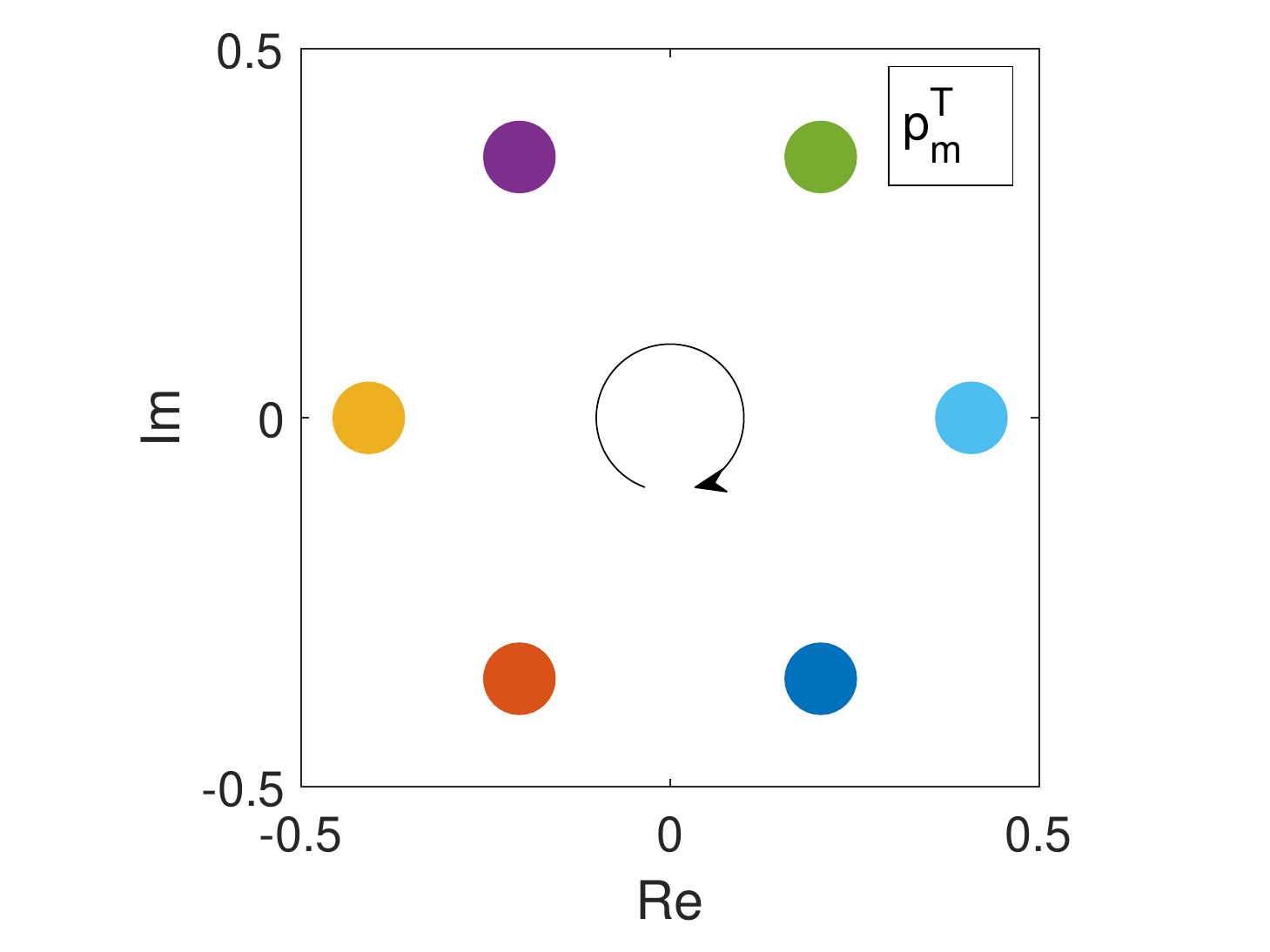} &  \includegraphics[height=3.0 cm, valign=t]{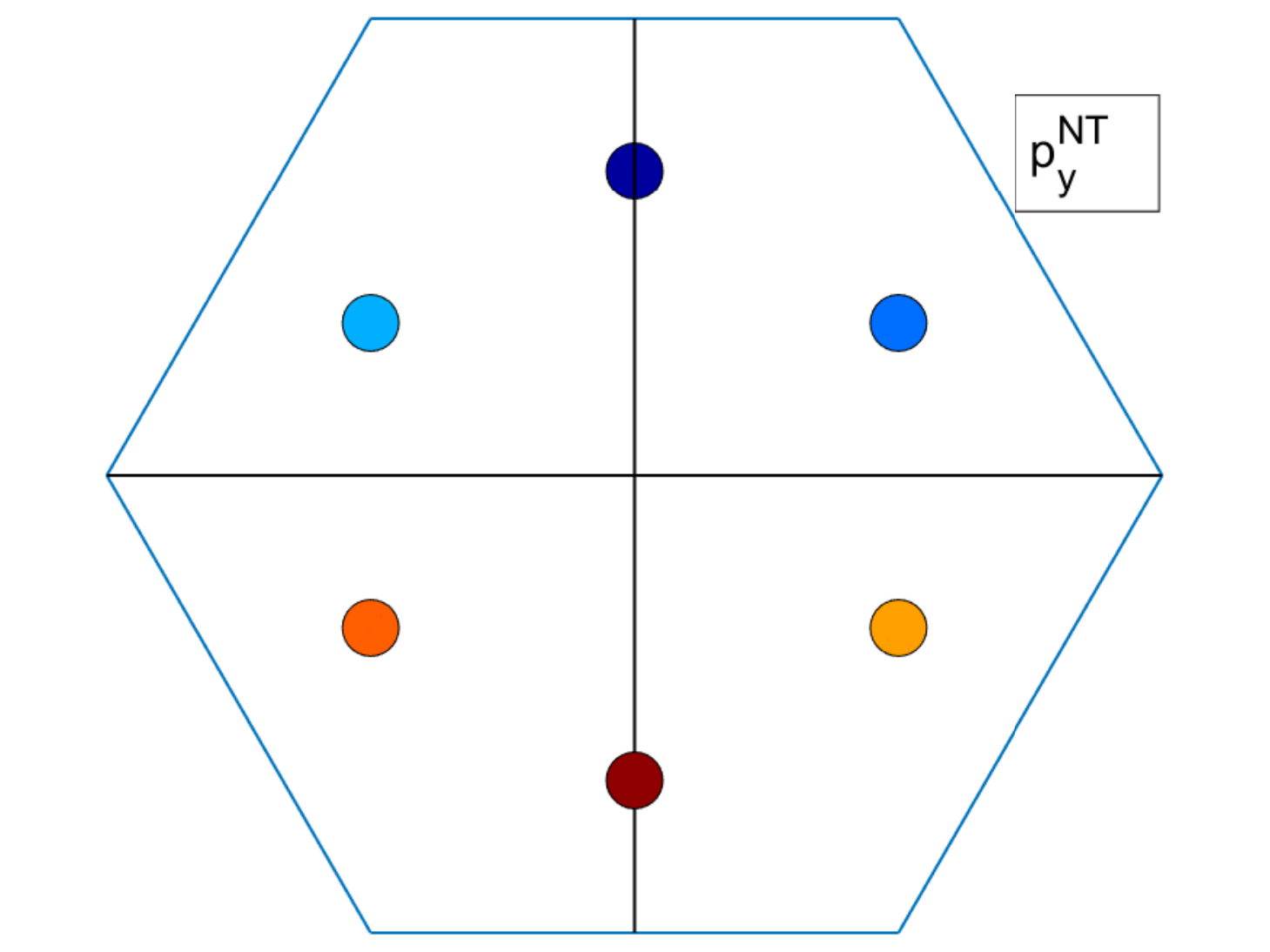} & \includegraphics[height=3.2 cm, valign=t]{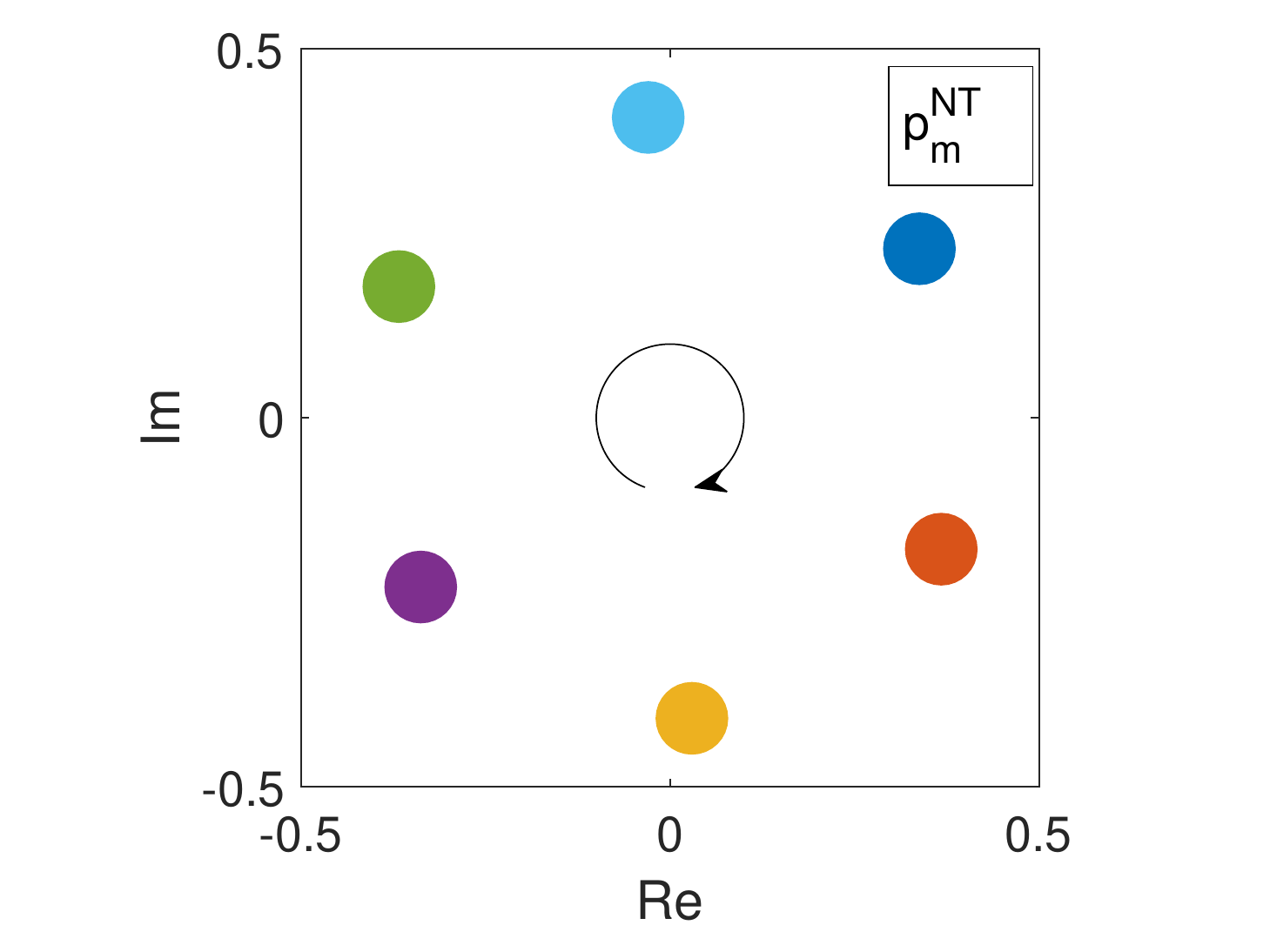} 
\end{tabular}
\caption{Eigenmodes of the infinite-periodic metamaterial. (a),(c) Dipole modes $p_x$,$p_y$ and quadrupole modes $d_{xy}$,$d_{x^2-y^2}$ of the trivial (non-trivial) topological regime corresponding to the $\Gamma$ point frequencies in Fig. \ref{QSHE_infinite}(b) (Fig. \ref{QSHE_infinite}(c)). Plotted is the out-of-plane displacement magnitude of each mass in the unit-cell. (b),(d) The associated pseudo-spin modes $d_{\pm}$, labeled $d_p,d_m$, and $p_{\pm}$, labeled $p_p,p_m$, of the trivial (non-trivial) regime. Each mode is plotted in the complex plane, portraying the relative phase of the masses out-of-plane displacement. Counterclockwise (clockwise) arrow indicates pseudo-spin up (down) polarization.}
\label{QSHE_infinite_modes}
\end{center}
\end{figure}
The eigenstates corresponding to the $d$ and $p$ frequencies are depicted in Fig. \ref{QSHE_infinite_modes}(a),(b) for $\beta>\frac{1}{3}$, and in Fig. \ref{QSHE_infinite_modes}(c),(d) for $\beta<\frac{1}{3}$. In Fig. \ref{QSHE_infinite_modes}(a) and (c) the actual states are depicted, indicating the masses vibration amplitudes. We denote these states by quadrupoles $d_{xy}$ and $d_{x^2-y^2}$, and dipoles $p_{x}$ and $p_{y}$, as their symmetries are analogous to the electronic orbital shapes. In particular, $d_{xy}$ ($d_{x^2-y^2}$) is odd (even) symmetric about both the $x$ and $y$ axes, whereas $p_x$ ($p_y$) is even (odd) symmetric about the $x$ ($y$) axis and odd symmetric about the $y$ ($x$) axis.
The linear combinations $d_{\pm}=d_{x^2-y^2}\pm id_{xy}$ and $p_{\pm}=p_x\pm ip_y$ constitute pseudospins of the system, and are illustrated in Fig. \ref{QSHE_infinite_modes}(b) and (d). The $+$ ($-$) sign indicates pseudospin up (down) and is denoted by a counterclockwise (clockwise) arrow. 
Remarkably, the pseudospin multipolar states in our system result from a phase difference in the out-of-plane vibration of the six masses in the unit cell, without any physical rotation. \\
The parameter $\beta$ reflects on the coupling strength between the adjacent unit cells, indicating the pseudospin – orbit coupling. 
The transition between the $\beta>\frac{1}{3}$ and the $\beta<\frac{1}{3}$ regimes occurs through a gap closing, and imposes a band inversion. For $\beta>\frac{1}{3}$ the dipole states correspond to the lower bands, and the quadrupole states to the upper bands. Here $t_2<t_1$ and the pseudospin - orbit coupling is weak, rendering this regime as topologically trivial. By gradually increasing the stiffness of the $t_2$ springs (via the feedback operation) and thereby the pseudospin - orbit coupling strength, the dipole states overtake the quadrupole states, resulting in a transition to a topologically non-trivial regime, manifested by an inverted band-structure.
This phenomenon is equivalent to the inversion of the conduction and the valence bands due to spin - orbit coupling in electronic topological insulators.
The topological character of the transition between regimes is also manifested by the spin Chern number [31], and can be shown to obtain non-trivial values in the $\beta<\frac{1}{3}$ case.
Consequently, the eigenstates are labeled by the superscript $T$ in Fig. \ref{QSHE_infinite_modes}(a),(b), and $NT$ in Fig. \ref{QSHE_infinite_modes}(c),(d).

\begin{figure}[t] 
\begin{center}
\begin{tabular}{l l l l l l l l}
\small{\textbf{(a)}} & &  \small{\textbf{(b)}} & & \small{\textbf{(c)} Positive $v_g$} & & \small{\textbf{(d)} Negative $v_g$} &  \\
\includegraphics[height=5.5 cm, valign=c]{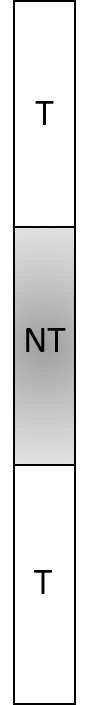} & & \includegraphics[height=6.0 cm, valign=c]{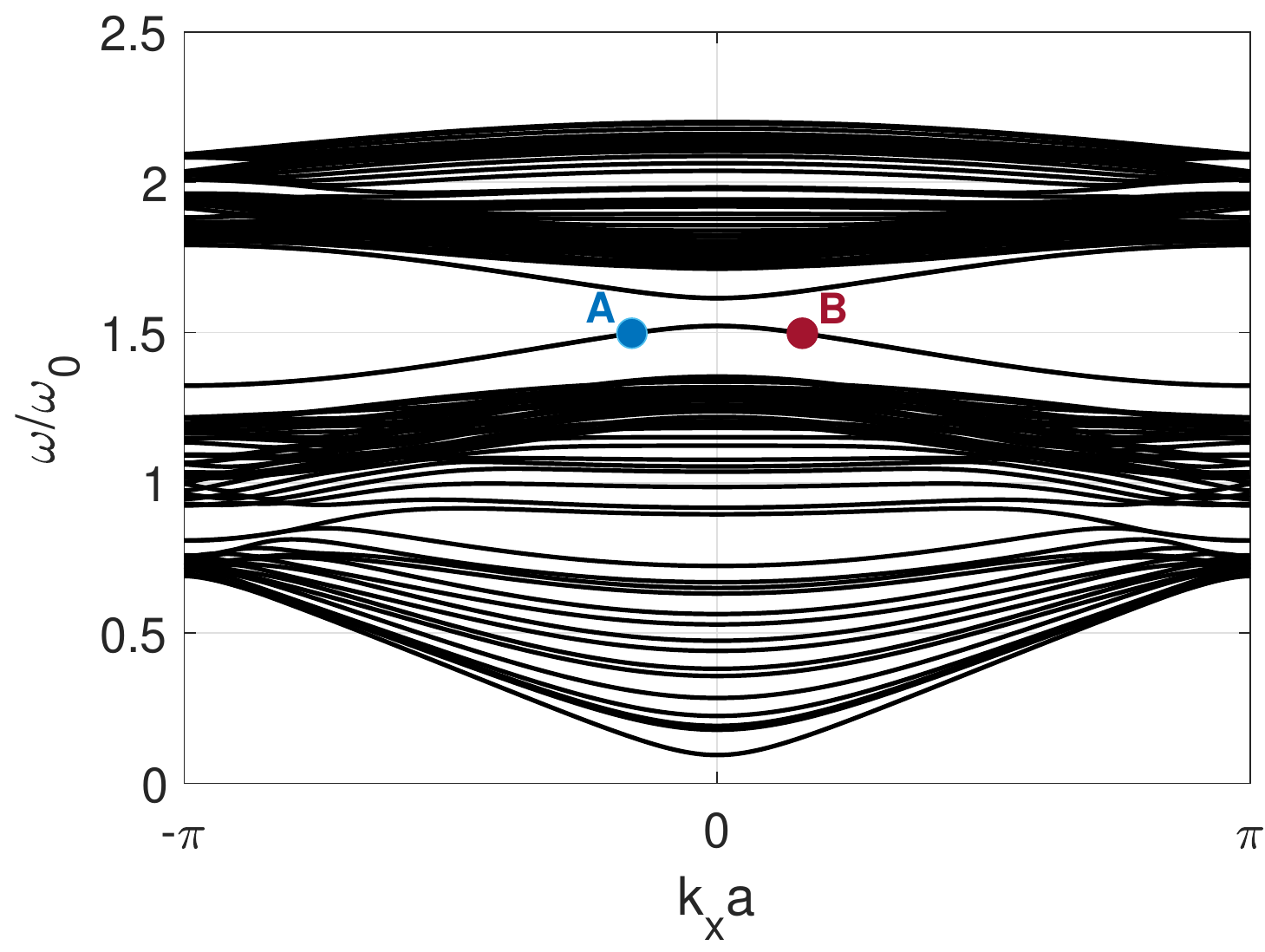} & & \includegraphics[height=6.0 cm, valign=c]{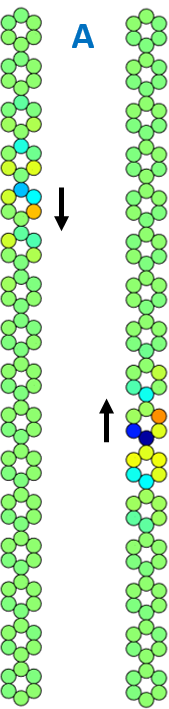}  & & \includegraphics[height=6.0 cm, valign=c]{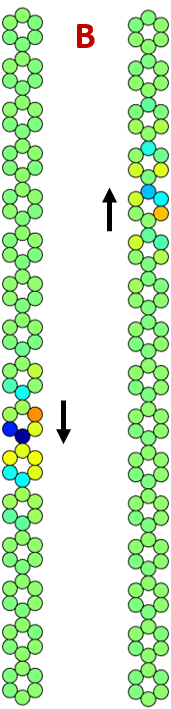}   &  \includegraphics[height=3.5 cm, valign=c]{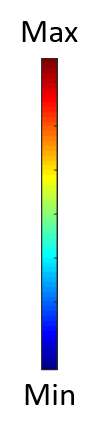}
\end{tabular}
\caption{Finite system analysis. (a) Super-cell schematic. (b) Dispersion profile of $6$ non-trivial honeycomb cells cladded by $5$ trivial cells on each side. Two degenerate pairs of interface states emerge in the bulk band-gap. (c),(d) out-of-plane displacement modes of points $A$ and $B$, corresponding to forward and backward transmission channels (positive and negative group velocity $v_g$). The modes are confined to the interfaces and carry an opposite counterclockwise and clockwise polarization of the out-of-plane displacement field (up and down arrows).}
\label{QSHE_interface}
\end{center}
\end{figure}

\pagebreak

One of the key manifestations of classical analogies of topological insulators is the pseudospin-dependent uni-directional transmission of waves, which are localized at an interface of topologically trivial and non-trivial lattices, with the bulk remaining insulating.
We therefore analyze the wave behavior of adjacently connected topologically distinct lattices.
We construct a super-cell in form of a strip in the $\bf{y}$ direction, consisting of $6$ topologically non-trivial cells ($\beta=1.2\cdot\frac{1}{3}$), cladded by $5$ topologically trivial cells ($\beta=0.7\cdot\frac{1}{3}$) at each end. The super-cell is schematically illustrated in Fig. \ref{QSHE_interface}(a). 
In the $\bf{x}$ direction a periodic boundary condition is assumed. 
The resulting dispersion profile is depicted in Fig. \ref{QSHE_interface}(b). 
Compared to the infinite lattice dispersion profile in Figs. \ref{QSHE_infinite}(b),(c), we obtain two time-reversed pairs of band-gap crossing states, indicated by the double degenerated curves. Each pair corresponds to one of the two interfaces in the super-cell.   
For each interface, the two states have opposite group velocities and an opposite polarization, corresponding to clockwise (pseudo-spin down) and counterclockwise (pseudo-spin up) rotating modes, localized at the interface. \\
The polarization is exhibited through the harmonic evolution of these modes. 
For example, we consider the points $A$ and $B$ in Fig. \ref{QSHE_interface}(b), associated with the normalized frequency $\Omega=1.5$. The eigenstates corresponding to $A$ and $B$ are respectively plotted in Fig. \ref{QSHE_interface}(c) and Fig. \ref{QSHE_interface}(d). The confinement to the interfaces is clearly observed. 
We thus obtain four spatially separated transmission channels.
The top interface contains a forward channel (positive group velocity $v_g>0$, the $A$ point) with pseudo-spin down, and a backward channel (negative group velocity $v_g<0$, the $B$ point) with pseudo-spin up, and vice-versa for the bottom interface.
The implication is that exciting the metamaterial with a source of frequency inside the bulk band-gap, at the vicinity of an interface, will induce forward and backward propagating waves of opposite polarization. 
Inducing a strictly uni-directional wave is possible via a phased excitation, aligned with the appropriate polarization. 
We note that the interface modes carry a topological character, since they emerged due to difference in topology of the concatenated lattices. However, as $C_6$ symmetry is not preserved at the interface, these modes are not completely topologically protected.
This problem may be resolved by replacing the abrupt interface with a gradual one, thus locally preserving $C_6$ symmetry of the cells.

Finally, we demonstrate that the metamaterial in Fig. \ref{QSHE_scheme}(a), which was created by the control algorithm in \eqref{eq:Motion_OL}, indeed supports topologically protected wave propagation.
We perform dynamical simulations of the same host structure as in the main text, a $20\times 40$ honeycomb lattice with six masses in each cell. 
The controllers are programmed in a way that the lattice is artificially divided by a $V$-shaped interface into a trivial and a non-trivial zone with $\beta=0.7\cdot\frac{1}{3}$ and $\beta=1.2\cdot\frac{1}{3}$, respectively. We stress that the artificial interface shape and the spring constants values are determined exclusively by the control program, and are created in real-time on top of the nominal lattice with all identical springs and masses. 
The nominal lattice parameters are $a_0=0.05\sqrt{3}$ and $t_0\approx 1.4\cdot 10^5$, with an addition of a small viscous damping $\zeta=0.01$. We excite the structure on the interface, at the location indicated by the blue arrow in Figs. \ref{QSHE_simulations}(a) and \ref{QSHE_simulations}(b), by an harmonic force with the frequency $\omega/\omega_0=1.5$. This frequency corresponds to the edge states in Fig. \ref{QSHE_interface}(b). \\
The time response is depicted in Figs. \ref{QSHE_simulations}(a) and \ref{QSHE_simulations}(b) at time instances $T_1$ and $T_2>T_1$, respectively. 
Since the harmonic force excites simultaneously both pseudo-spin-up and pseudo-spin-down modes (unlike a selective phased excitation), we observe wave propagation in both directions along the interface.
Due to topological protection, the wave traverses the sharp corner of the $V$ shape, and remains localized on the interface without scattering into the bulk.

\begin{figure}[h] 
\begin{center}
\begin{tabular}{l l l l}
\small{\textbf{(a)} $T_1$} &  \includegraphics[height=5.5 cm, valign=t]{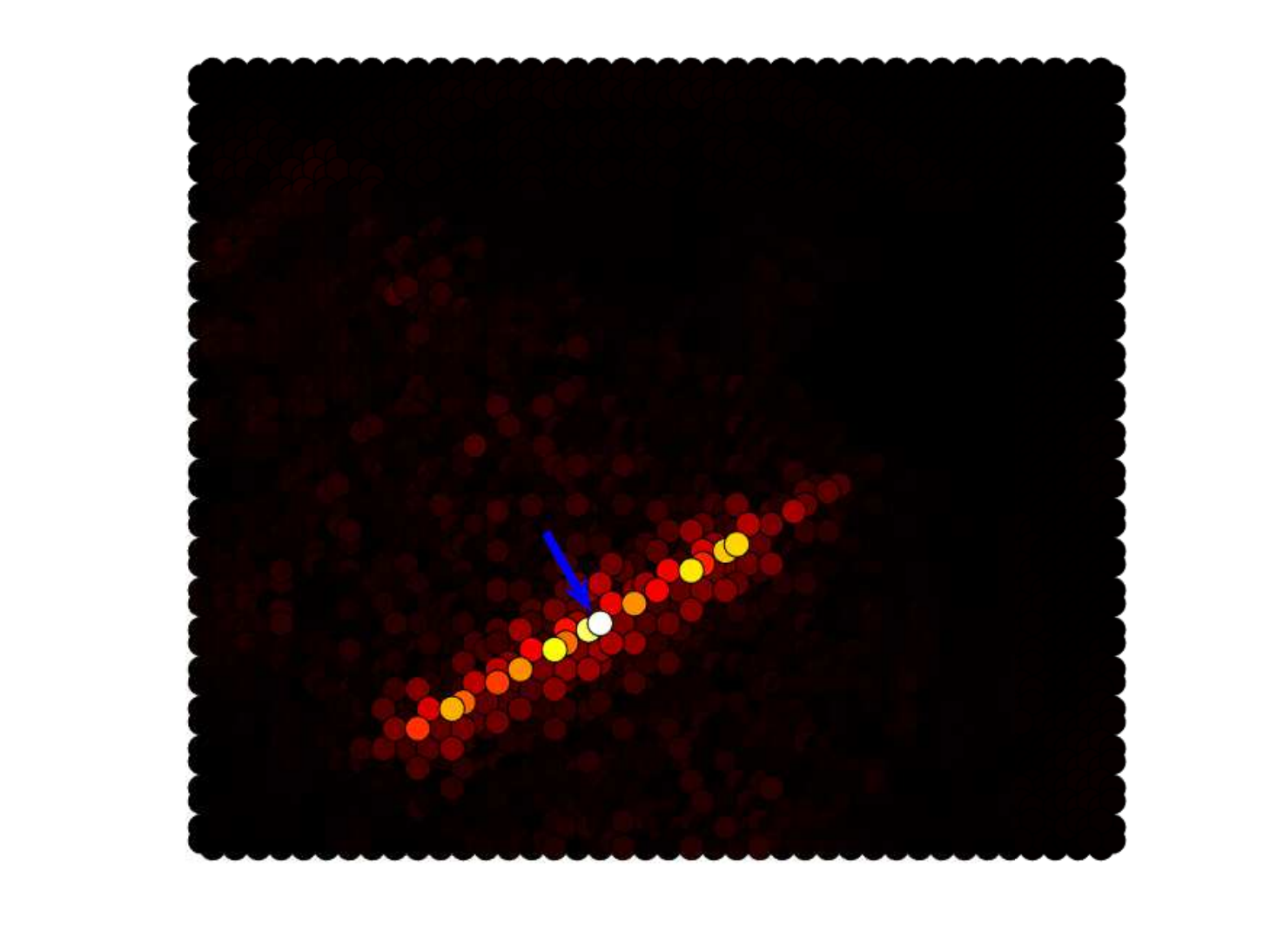} & \small{\textbf{(b)} $T_2$} & \includegraphics[height=5.5 cm, valign=t]{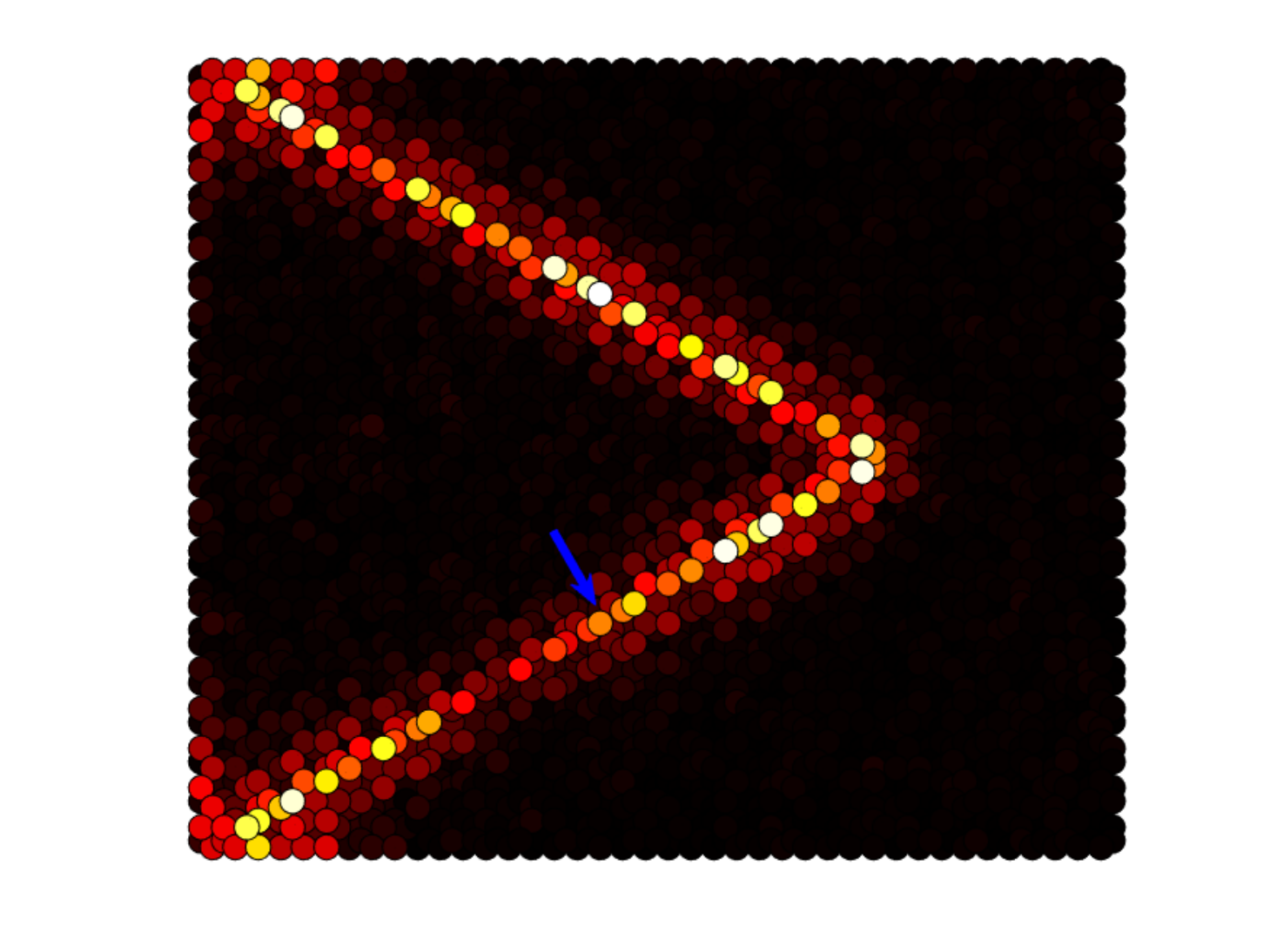}
\end{tabular}\\
\includegraphics[height=0.7 cm, valign=t]{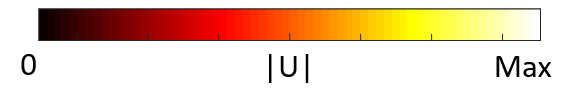}
\caption{Dynamical simulations of a $20\times 40$ honeycomb cells metamaterial with a $V$-shaped artificial interface. The source is of frequency $\omega/\omega_0=1.5$, located on the interface (blue arrow). Snapshots of the time response at $T_1$ (a) and $T_2$ (b). Wave propagation is localized on the interface, manifesting the topological phenomenon.}
\label{QSHE_simulations}
\end{center}
\end{figure}

\end{widetext}

\end{document}